%% file: Journal__TIT_v1.tex
\documentclass[journal,11pt,onecolumn,draftclsnofoot]{IEEEtran}
\usepackage{cite}
\usepackage{graphicx,color,epsfig,rotating}
\usepackage{amsfonts,amsmath,amssymb,bbm}
\usepackage{setspace}
\usepackage{algorithm}
\usepackage[algo2e]{algorithm2e} 
\usepackage{subcaption}
\usepackage{mdwtab}
\usepackage{placeins}
\usepackage{psfrag, graphicx}
\usepackage[latin1]{inputenc}
\usepackage{multirow}
\usepackage{stfloats}
\usepackage{tabularx} 
\usepackage{booktabs} 
\usepackage{url}
\usepackage{bm}
\usepackage{geometry}
\usepackage{pstricks}
 \allowdisplaybreaks

\input{macros}

\begin{document}
\title{Optimal Rate Region for Multi-server Secure Aggregation with User Collusion}
\input{author_TIT.tex}

\maketitle
\IEEEpeerreviewmaketitle

\input{sections_v1/abstract.tex}


\section{Introduction}
\label{sec:intro}

Federated learning (FL) enables collaborative model training over distributed users while keeping raw data local, and has become a key paradigm for large-scale and privacy-sensitive learning systems~\cite{konecny2016federated,kairouz2021advances,mcmahan2017communication,yang2018applied}. In a typical FL workflow~\cite{beltran2023decentralized}, users compute local updates and transmit them to an aggregator, which combines these updates to form a global model. Although raw data are not directly shared, it is now well understood that intermediate model updates may still leak sensitive information about individual users through inference attacks such as model inversion and gradient leakage. As a result, keeping data local alone is insufficient to guarantee privacy in FL systems.

Secure aggregation (SA) has therefore been introduced as a fundamental mechanism for privacy-preserving FL. Its goal is to ensure that the aggregator can recover only the aggregate of users' updates while learning nothing about any individual contribution. Early protocols, including the seminal work of Bonawitz \emph{et al.},~\cite{bonawitz2017practical, bonawitz2016practical} demonstrated that this objective can be achieved against an honest-but-curious server. Subsequent works have further improved the robustness and efficiency of SA schemes by addressing practical challenges such as user dropouts, communication overhead, and scalability, establishing secure aggregation as a core building block of modern FL systems.

Beyond protocol design, secure aggregation has also been studied from an information-theoretic perspective~\cite{9834981}, with the aim of characterizing its fundamental performance limits under perfect security guarantees. In this framework, security is defined in the strongest sense: the aggregator is required to learn nothing beyond the desired aggregate, regardless of its computational power. A central question is to determine the minimum communication and key rates needed to simultaneously satisfy correctness and perfect security. In the single-server setting, Zhao and Sun~\cite{zhao2023secure} provided the first complete characterization of the optimal rate region, showing that the minimum user-to-server communication rate and individual key rate are both one symbol per input symbol, while the optimal source key rate scales linearly with the number of users. This line of work has subsequently been extended to a variety of more general settings, including user dropout~\cite{so2022lightsecagg}, collusion~\cite{jahani2022swiftagg,jahani2023swiftagg+,li2023weakly,li2025weakly}, user selection~\cite{zhao2022mds,zhao2023optimal}, groupwise keys~\cite{zhao2023secure,wan2024information,wan2024capacity}, oblivious servers~\cite{sun2023secure}, decentralized secure aggregation~\cite{Zhang_Li_Wan_DSA,Li_Zhang_GroupwiseDSA,Li_Zhang_WeaklyDSA}, and hierarchical secure aggregation~\cite{zhang2024optimal,10806947,egger2024privateaggregationhierarchicalwireless,zhang2025fundamental,11195652,li2025collusionresilienthierarchicalsecureaggregation,egger2023private,lu2024capacity,Li_Zhang_WeaklyHSA}. These extensions provide deeper insights into the fundamental tradeoffs among security guarantees, communication efficiency, and key randomness.

Motivated by recent advances in hierarchical and decentralized secure aggregation~\cite{zhang2024optimal,Zhang_Li_Wan_DSA}, this paper studies secure aggregation in a multi-server setting. In such systems, multiple aggregation servers jointly participate in the computation, with each server connected to a group of users. Compared to the typical single-server model, the multi-server architecture is appealing from a practical standpoint, as it improves scalability, fault tolerance, and resilience to server compromise. However, it also introduces new fundamental challenges that are absent in existing secure aggregation models. In particular, each server simultaneously receives two distinct types of information: the messages sent by its associated users and the messages exchanged with other servers during the aggregation process. This joint observation of user-level and server-level information does not arise in conventional single-server secure aggregation, nor in existing decentralized or hierarchical models. At the same time, all servers are required to recover the global sum of all users' inputs, even though each server directly observes only a subset of the user messages. This requirement creates a nontrivial coupling between correctness and security constraints that fundamentally differentiates the multi-server setting from prior models.

The problem becomes more intricate in the presence of user collusion, where a subset of compromised users may reveal their inputs and individual keys to aggregation servers. In a multi-server architecture, collusion can occur in two forms: users directly connected to a server, and users connected to other servers, both of whose information becomes directly available once they collude. When combined with the inter-server messages inherent to the multi-server design, these two types of collusion can significantly amplify information leakage by creating new dependencies across servers. The key challenge is therefore to ensure perfect security against such collusion while balancing communication efficiency and key randomness across different layers of the system.

In this work, we study multi-server secure aggregation with user collusion from an information-theoretic perspective. We consider a two-hop network consisting of multiple aggregation servers and multiple users per server, where users communicate only with their associated server and servers exchange messages over a fully connected network. We adopt an honest-but-curious threat model and allow each server to collude with a bounded number of users. Under this model, we derive the complete optimal rate region, characterizing the minimum user-to-server communication rate, server-to-server communication rate, individual key rate, and source key rate required to achieve both correctness and perfect security. Our results provide a precise information-theoretic understanding of the fundamental limits of secure aggregation in multi-server systems and offer design guidelines for communication- and key-efficient secure aggregation schemes in large-scale distributed learning networks.

\subsection{Summary of Contributions}

The main contributions of this paper are summarized as follows:
\begin{itemize}
    \item We introduce an information-theoretic model for multi-server secure aggregation with user collusion, where multiple aggregation servers jointly compute the sum of users' inputs, each user is associated with a single server, and collusion is modeled by allowing servers to access the inputs and keys of a subset of users.

    \item We propose a communication- and key-efficient secure aggregation scheme that achieves perfect correctness and information-theoretic security under the considered collusion model, enabling all servers to simultaneously recover the global input sum.

    \item We establish matching converse bounds and fully characterize the optimal rate region of multi-server secure aggregation, identifying the minimum user-to-server communication rate, server-to-server communication rate, individual key rate, and source key rate required for secure aggregation in the presence of user collusion.
\end{itemize}

\emph{Notation.}
For integers $m \le n$, let $[m:n] \triangleq \{m,m+1,\ldots,n\}$, and write $[1:n]$ simply as $[n]$.
Calligraphic letters (e.g., $\mathcal{A}, \mathcal{B}$) denote sets.
The Cartesian product of $\mathcal{A}$ and $\mathcal{B}$ is defined as
$\mathcal{A} \times \mathcal{B} \triangleq \{(a,b): a \in \mathcal{A}, b \in \mathcal{B}\}$.
For a collection $\{A_i\}_{i \in [n]}$, we write $\{A_1,\ldots,A_n\}$, and for an index set $\mathcal{I} \subseteq [n]$, define
$A_{\mathcal{I}} \triangleq \{A_i\}_{i \in \mathcal{I}}$.
Finally, the set difference is denoted by
$\mathcal{A} \backslash \mathcal{B} \triangleq \{x \in \mathcal{A}: x \notin \mathcal{B}\}$.

\section{Problem Formulation}
\label{sec: problem description}
\begin{figure}[h]
    \centering
    \includegraphics[width=0.48\textwidth]{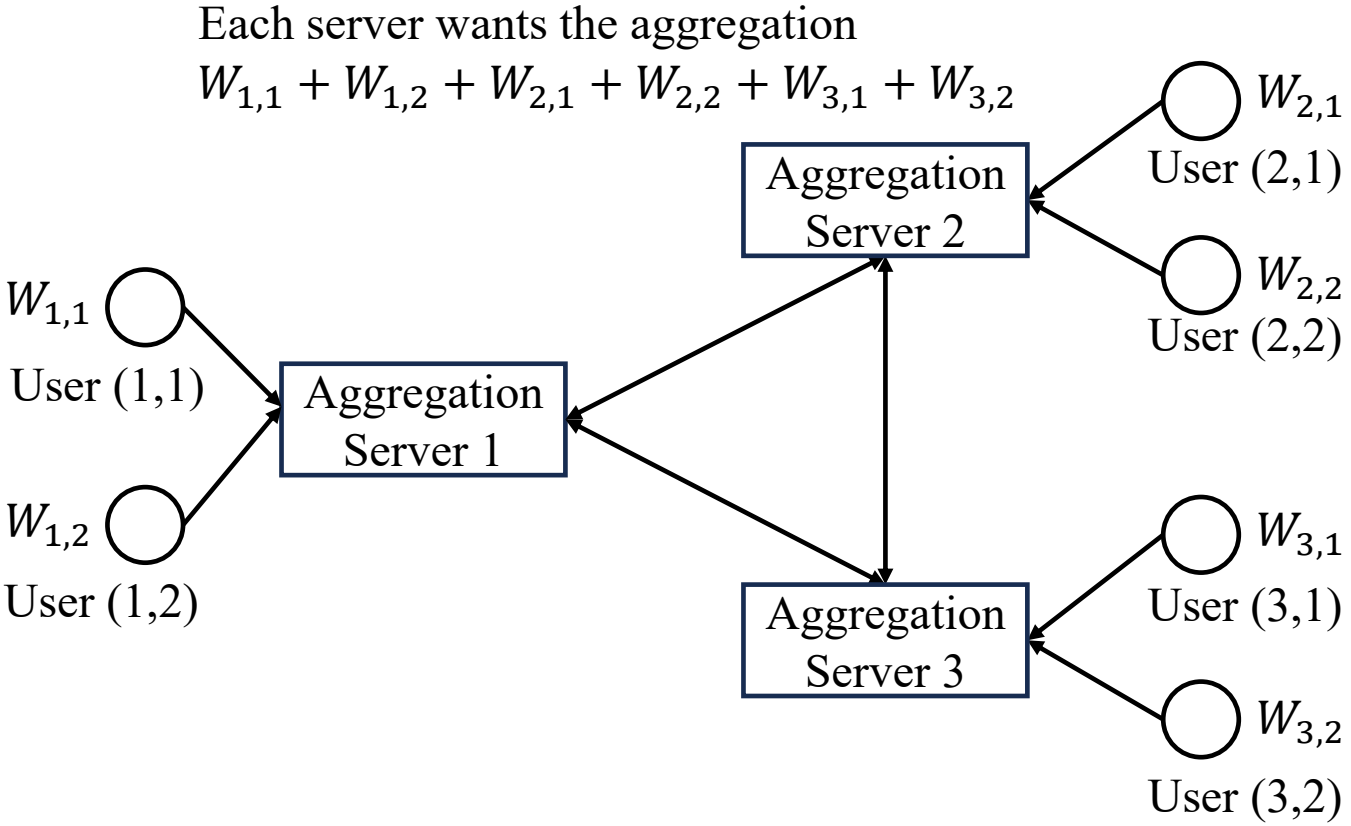}
    \caption{\small  Multi-server Secure Aggregation with $U=3$ servers and $V=2$ users. Each Aggregation server aims to compute the sum of inputs $W_{1,1}+W_{1,2}+W_{2,1}+W_{2,2}+W_{3,1}+W_{3,2}$ of all users.}
    \label{fig:model}
\end{figure}

We consider the \secagg problem in a two-hop communication network comprising $U \geq 3$ servers and $UV$ users in total. Each server is connected to a distinct group of \(V\) users through error-free channels in the first hop, and the servers are fully interconnected via error-free broadcast channels in the second hop, as depicted in Fig.~\ref{fig:model} (An example with $U=3, V=2$). 
We denote the $v^{\rm th}$ user of the $u^{\rm th}$ server as $(u,v)\in [U]\times [V]$.  Let $\mathcal{M}_u \triangleq \{(u,v)\}_{v\in [V]}$ denote the set of users associated with server $u$. Each user $(u,v)$ possesses an \emph{input} $W_{u,v}$ consisting of $L$ independent and identically distributed (i.i.d.) symbols uniformly drawn from a finite field $\mathbb{F}_q$. In addition, each user holds a secret key $Z_{u,v}$, referred to as an \emph{individual key}, which contains $L_Z$ symbols over $\mathbb{F}_q$.
The inputs $W_{[U]\times [V]} \eqdef \{W_{u,v}\}_{(u,v)\in [U]\times [V]}$ are assumed to be independent of the keys $Z_{[U]\times [V]}\eqdef  \{Z_{u,v}\}_{(u,v)\in [U]\times [V]}$; that is,
\begin{eqnarray}
   &&H\left(Z_{[U]\times [V]},W_{[U]\times [V]}\right) = H\left(Z_{[U]\times [V]}\right)+ \sum_{(u,v)\in[V]\times [U]} H(W_{u,v}), \label{ind} \\
   && H(W_{u,v}) = L ~(\mbox{in $q$-ary units}), ~\forall~ (u,v)\in [U]\times [V]. \label{h2}
\end{eqnarray}

The individual keys $Z_{[U]\times [V]}$, can be arbitrarily correlated and are generated from a \emph{source key} variable $\zsigma$, which consists of $\lzsigma$ symbols from a finite field $\mathbb{F}_q$, such that
\begin{align}
    H\left(Z_{[U]\times [V]}|Z_{\Sigma}\right)=0. \label{sourcekey}
\end{align}
All randomness required for key construction is encapsulated in the source key $\zsigma$, which consists of mutually independent symbols. A trusted authority performs key assignment offline so that every user $(u,v)$ possesses its designated key $Z_{u,v}$ in advance of the aggregation task.

The system follows a two-hop communication model. During the first hop, each user $(u,v)$ sends an $L_X$-symbol message $X_{u,v}$ to server $u$. This message is deterministically produced from the user's input $W_{u,v}$ together with its private key $Z_{u,v}$. Accordingly, the relationship can be expressed as
\begin{align}
    H(X_{u,v}|W_{u,v}, Z_{u,v})=0, \forall (u,v)\in[U]\times[V]. \label{messageX}
\end{align}
In the second hop, each server $u$ forms an $L_Y$-symbol broadcast message $Y_u$ and delivers it to the remaining $U-1$ servers. The message $Y_u$ is deterministically generated from the collection of messages received from its associated users, namely $\{X_{u,v}\}_{v\in [V]}$. This relationship can be described as
\begin{align}
    H(Y_u|\{X_{u,v}\}_{v\in[V]})=0,\forall u\in[U].\label{messageY}
\end{align}

Based on the messages from its associated users and those exchanged among the other servers, server $k\in[U]$ is required to recover the global sum of inputs, i.e.,
\be
\label{correctness}
\mbox{[Correctness]}~H\left(\sum_{(u,v)\in[V]\times [U]}W_{u,v}\Bigg| \{Y_u\}_{u\in[U]\setminus \{k\}},\{X_{k,v}\}_{v\in [V]}\right) =0, \forall k\in [U]. 
\ee

The security requirement ensures that each server gains no information about the users' inputs $W_{[U]\times [V]}$ beyond the intended aggregated result $\sum_{(u,v)\in[U]\times[V]} W_{u,v}$, even when it colludes with any subset $\Tc$ of at most $T$ users, i.e., $|\mathcal{T}|\leq T$.
Formally, this constraint can be expressed in terms of mutual information as
\begin{align}
\label{security}
&\mbox{[Security]}~I\left(\{Y_u\}_{u\in[U]\setminus \{k\}},\{X_{k,v}\}_{v\in [V]}; W_{[U]\times [V]}
\bigg|\sum_{(u,v)\in[V]\times [U]}W_{u,v}, \{W_{u,v}, Z_{u,v } \}_{(u,v )\in \Tc} \right)=0, \notag\\
&~~~~~~~~~~~~~~~~~~~~~~~~~~~~~~~~~~~~~~~~~~~~~~~~~~~~~~~~~~~~~~~~~~~~~~~~\forall k\in[U], \Tc\subseteq [U]\times [V] : |\Tc|\le T. 
\end{align}

The communication rates $R_X$ and $R_Y$ specify the number of symbols contained in each message $X_{u,v}$ and $Y_u$, respectively, per input symbol. Similarly, the individual key rate $R_Z$ and the source key rate $R_{Z_{\Sigma}}$ indicate the number of symbols contained in each individual key $Z_{u,v}$ and the source key $Z_{\Sigma}$, respectively, per input symbol, i.e.,
\be 
\label{eq: def of rate}
R_X \eqdef \frac{L_X}{L}, R_Y \eqdef \frac{L_Y}{L},
R_Z \eqdef  \frac{L_Z}{L}, R_{Z_{\Sigma}} \eqdef  \frac{L_{Z_{\Sigma}}}{L}.
\ee 
A rate tuple $(R_X, R_Y, R_Z, R_{Z_{\Sigma}})$ is said to be achievable if there exists a secure coding strategy, consisting of the key generation and message construction procedures for $X_{u,v}$, $Y_u$, $Z_{u,v}$, and $Z_{\Sigma}$,  that satisfy the structural conditions (\ref{messageX}) and (\ref{messageY}), achieve the corresponding communication and key rates $R_X$, $R_Y$, $R_Z$, and $R_{Z_{\Sigma}}$, and ensure both the correctness condition (\ref{correctness}) and the security condition (\ref{security}).
The optimal rate region $\Rc^*$ is defined as the closure of all achievable rate tuples.


\section{Main Result}
\label{sec: main result}
\begin{theorem}
\label{thm1}
For the multi-server secure aggregation problem with $U \ge 3$ servers, $V$ users per server, and at most $T$ colluding users, the optimal rate region is given by 
\begin{eqnarray}
\mathcal{R}^*=\{(R_X,R_Y,R_Z,R_{Z_\Sigma}): R_X\geq 1, R_Y\geq 1, R_Z\geq 1, R_{Z_{\Sigma}}\geq \min \{U+V+T-2, UV-1\}\}
\end{eqnarray}
\end{theorem}

Theorem~1 characterizes the optimal communication and key rate region for the multi-server secure aggregation problem with user collusion. The result shows that the user-to-server and server-to-server communication rates must satisfy $R_X \ge 1$ and $R_Y \ge 1$, and the individual key rate must satisfy $R_Z \ge 1$. In addition, the minimum source key rate is given by $R_{Z_\Sigma}^* = \min\{U+V+T-2,\; UV-1\}.$

The necessity of the communication rate constraints $R_X \ge 1$ and $R_Y \ge 1$ follows from the correctness requirement. Since each server is required to reconstruct the global sum of all users' inputs, the information carried by every individual input symbol must be conveyed through the two-hop communication links. Consequently, each user-to-server message and each server-to-server message must contain at least as much information as one input symbol.

The individual key rate constraint \(R_Z \ge 1\) is required by the security requirement. Each user's input must remain perfectly concealed from any server beyond the revealed global sum. In particular, even if all other users' inputs and keys are known, the uncertainty of a single user's input must be protected by its own key. Therefore, the individual key must contribute at least one symbol of independent randomness per input symbol, which implies \(R_Z \ge 1\).

The source key rate constraint captures the fundamental limitation imposed by security against colluding users. From the perspective of a given server, it observes $V$ masked messages from its associated users and $U-1$ aggregated messages from the other servers. Among these observations, one linear combination corresponds to the desired global sum and is intentionally revealed by the protocol. Conditioned on this sum, the remaining observations span at most $U+V-2$ independent linear dimensions. Moreover, when the server colludes with up to $T$ users, the colluding users may reveal up to $T$ additional input symbols together with their corresponding keys, which increases the amount of side information available to the server.

To ensure information-theoretic security, all remaining observable dimensions must be protected by independent randomness. Therefore, at least $U+V+T-2$ independent key symbols are required. On the other hand, since there are $UV$ individual keys in total and one linear dependence among them is necessary to allow recovery of the global sum, the source key rate cannot exceed $UV-1$. Taking the minimum of these two quantities yields the optimal source key rate $R_{Z_\Sigma}^*$.

\section{Achievability Proof of Theorem \ref{thm1}}
\label{sec:genenralscheme}
Before presenting the general achievability scheme for Theorem~\ref{thm1},
we first present two examples that illustrate the main ideas underlying the proposed multi-server secure aggregation design.
The general construction will be given subsequently.

\subsection{Achievability Proof of Example~\ref{example: U=3, V=2, T=0}}
\begin{example}
\label{example: U=3, V=2, T=0}
Consider the setting $(U,V,T)=(3,2,0)$, as illustrated in Fig.~\ref{fig:model}, where no users collude, i.e., $T=0$.
By Theorem~\ref{thm1}, the length of the source key is $L_{Z_{\Sigma}}=\min \{U+V+T-2,UV-1\}=3$.
Each input $W_{u,v}$ consists of one symbol over $\mathbb{F}_{11}$.
The source key $Z_{\Sigma}=(N_1,N_2,N_3)$ comprises three \iid uniformly distributed random variables over $\mathbb{F}_{11}$.
The individual keys are chosen as follows.
\begin{align}
\label{eq: indiv keys, example 1}
& Z_{1,1} = N_1,\; Z_{1,2}=N_2,\; Z_{2,1}=N_3,\; Z_{2,2} = N_1+2N_2+3N_3,\notag\\
&  Z_{3,1}=N_1+3N_2+4N_3,\; Z_{3,2}=-(3N_1+6N_2+8N_3).
\end{align}

Each user $(u,v)$ sends a \msg  $X_{u,v}=W_{u,v}+Z_{u,v}$ to server $u$. In particular,
\begin{align}
\label{eq: X1,X2,X3, example 1}
X_{1,1} &=W_{1,1} + N_1,~~X_{1,2} =W_{1,2} + N_2,~~X_{2,1} =W_{2,1} + N_3,~~X_{2,2} =W_{2,2} + N_1+2N_2+3N_3,\notag\\
X_{3,1} &=W_{3,1} + N_1+3N_2+4N_3,~~~~~~~~~~X_{3,2} =W_{3,2} -(3N_1+6N_2+8N_3).
\end{align}
Each server $ u$ broadcasts $Y_u = \sum_{v=1}^2 X_{u,v}$ to other servers. In particular,
\begin{align}
\label{eq: Y1, Y2, example 1}
Y_1 &=W_{1,1} +W_{1,2}+ N_1+N_2,\notag\\
Y_2 &=W_{2,1} +W_{2,2} +N_1+2N_2+4N_3,\notag\\
Y_3 &=W_{3,1} +W_{3,2}-(2N_1+3N_2+4N_3).
\end{align}
Since $L_X=L_Y=L_Z=1$ and $L_{Z_{\Sigma}}=3$, the corresponding rates are
$R_X=R_Y=R_Z=1$ and $R_{Z_{\Sigma}}=3$.
Under this construction, each server $k\in[3]$ is able to reconstruct the global aggregation of all users' inputs by appropriately combining the messages it observes.
In particular, server $k$ sums the messages received from its associated users,
namely $\{X_{k,v}\}_{v\in[2]}$, together with the server-to-server messages
$\{Y_u\}_{u\in[3]\setminus\{k\}}$ forwarded by the other servers.
This operation yields $\sum_{v=1}^{2} X_{k,v} + \sum_{u\in[3]\setminus\{k\}} Y_u
= \sum_{(u,v) \in [3]\times [2]} W_{u,v}.$
To illustrate, consider server $k=1$.
By construction, it computes $X_{1,1}+X_{1,2}+Y_2+Y_3=(W_{1,1}+N_1)+(W_{1,2}+N_2)  +(W_{2,1}+W_{2,2}+N_1+2N_2+4N_3) +(W_{3,1}+W_{3,2}-(2N_1+3N_2+4N_3))= \sum_{(u,v) \in [3]\times [2]} W_{u,v},$
where all key symbols cancel out by the key construction in (\ref{eq: indiv keys, example 1}).
Hence, the correctness requirement is satisfied.
The security argument is given next.

An important property of the key construction in (\ref{eq: indiv keys, example 1}) is that
\emph{any three out of the six keys are mutually independent}.
Although each server observes $V+(U-1)=4$ masked messages, one linear combination of these
observations corresponds to the desired aggregation $\sum_{(u,v) \in [3]\times [2]} W_{u,v}$, which is intentionally revealed by the protocol.
Consequently, conditioned on the desired aggregation $\sum_{(u,v) \in [3]\times [2]} W_{u,v}$, the remaining observations span at most three independent linear dimensions.
Since any three keys among $\{Z_{u,v}\}_{(u,v)\in[3]\times[2]}$ are mutually independent,
these three dimensions are fully protected by independent randomness.
As a result, conditioned on the sum $\sum_{(u,v) \in [3]\times [2]} W_{u,v}$, the joint distribution of the observations $\{X_{k,v}\}_{v\in [2]} \quad \text{and} \quad \{Y_u\}_{u\in[3]\setminus\{k\}}$ is independent of the individual inputs $\{W_{u,v}\}_{(u,v)\in [3]\times [2]}$.
Hence, apart from the required aggregate, the server cannot infer any information about the inputs. We next formalize the above intuition,
\begin{align}
& I\left(X_{1,1},X_{1,2}, Y_2,Y_3; \{W_{u,v}\}_{(u,v)\in[3]\times [2]}\Bigg|\sum_{(u,v)\in[3]\times [2]}W_{u,v}  \right)\notag\\
=&H\left(X_{1,1},X_{1,2}, Y_2,Y_3\Bigg|\sum_{(u,v)\in[3]\times [2]}W_{u,v}  \right)-H\left(X_{1,1},X_{1,2}, Y_2,Y_3\Bigg|\sum_{(u,v)\in[3]\times [2]}W_{u,v}, \{W_{u,v}\}_{(u,v)\in[3]\times [2]}  \right)\\
\overset{(\ref{eq: X1,X2,X3, example 1})(\ref{eq: Y1, Y2, example 1})}{=}&H\left(W_{1,1}+N_1,W_{1,2}+N_2, W_{2,1} +W_{2,2}+N_1+2N_2 +4N_3,W_{3,1} +W_{3,2}-(2N_1+3N_2+4N_3)\Big|\right.\\
&\left.\sum_{(u,v)\in[3]\times [2]}W_{u,v}  \right)-H\left(W_{1,1}+N_1,W_{1,2}+N_2, W_{2,1} +W_{2,2}+N_1+2N_2 +4N_3,\right.\notag\\
&\left.W_{3,1} +W_{3,2}-(2N_1+3N_2+4N_3)\Bigg|\sum_{(u,v)\in[3]\times [2]}W_{u,v}, \{W_{u,v}\}_{(u,v)\in[3]\times [2]}  \right)\\
=&H\left(W_{1,1}+N_1,W_{1,2}+N_2, W_{2,1} +W_{2,2}+N_1+2N_2 +4N_3,W_{3,1} +W_{3,2}-(2N_1+3N_2+4N_3),\right.\notag\\
&\left.\sum_{(u,v)\in[3]\times [2]}W_{u,v}  \right)-H\left(\sum_{(u,v)\in[3]\times [2]}W_{u,v}  \right)\notag\\
&-H\left(N_1,N_2, N_1+2N_2 +4N_3,-(2N_1+3N_2+4N_3)\Bigg|\sum_{(u,v)\in[3]\times [2]}W_{u,v}, \{W_{u,v}\}_{(u,v)\in[3]\times [2]}  \right)\\
\overset{(\ref{ind})(\ref{correctness})}{=}&H\left(W_{1,1}+N_1,W_{1,2}+N_2, W_{2,1} +W_{2,2}+N_1+2N_2 +4N_3,W_{3,1} +W_{3,2}-(2N_1+3N_2+4N_3)  \right)\notag\\
&-H\left(\sum_{(u,v)\in[3]\times [2]}W_{u,v}  \right)-H\left(N_1,N_2, N_3  \right)\label{eq:tt111}\\
=&4-1-3=0
\end{align}
The first term in (\ref{eq:tt111}) follows from the independence between the inputs and the keys, and $\sum_{(u,v)\in[3]\times[2]} W_{u,v}$ corresponds to the aggregate of the remaining messages.
Similarly, the third term in (\ref{eq:tt111}) also follows from the independence between the inputs and the keys, and $-(2N_1+3N_2+4N_3)$ is the linear combination of the remaining three keys.
\end{example}

In the following, we present a full-fledged example with $U=3$ servers, each associated with $V=3$ users, and with up to $T=2$ colluding users, to further illustrate the proposed design.
\subsection{Achievability Proof of Example \ref{example: U=3, V=3, T=2}}
\begin{example}
\label{example: U=3, V=3, T=2}
Consider the setting $(U,V,T)=(3,3,2)$, consisting of $U=3$ servers, each connected to $V=3$ users, and with each server allowed to collude with up to $T=2$ users.
By Theorem~\ref{thm1}, the length of the source key is $L_{Z_{\Sigma}}=\min \{U+V+T-2,UV-1\}=6$.
Each input $W_{u,v}$ contains one symbol from $\mathbb{F}_{17}$. 
The source key $Z_{\Sigma} =(N_1, N_2, N_3, N_4, N_5, N_6) $ contains 6 \iid uniform random variables from $\mathbb{F}_{17}$. The \indiv keys are chosen as 
\begin{align}
\label{eq: indiv keys, example 2}
& Z_{1,1} = N_1,\; Z_{1,2}=N_2,\;Z_{1,3}=N_3,\; Z_{2,1}=N_4,\; Z_{2,2} = N_5,\; Z_{2,3} = N_6,\notag\\
&  Z_{3,1}=N_1+2N_2+3N_3+4N_4+5N_5+6N_6,\; Z_{3,2}=N_1+3N_2+4N_3+5N_4+6N_5+7N_6,\notag\\
&Z_{3,3}=-(3N_1+6N_2+8N_3+10N_4+12N_5+14N_6).
\end{align}
Each user $(u,v)$ sends a \msg  $X_{u,v}=W_{u,v}+Z_{u,v}$ to server $u$, 
\begin{align}
\label{eq: X1,X2,X3, example 2}
X_{1,1} &=W_{1,1} + N_1,~~X_{1,2} =W_{1,2} + N_2,~~X_{1,3} =W_{1,3} + N_3,~~X_{2,1} =W_{2,1} + N_4,~~X_{2,2} =W_{2,2} + N_5,\notag\\
X_{2,3} &=W_{2,3} + N_6,~~X_{3,1} =W_{3,1} + N_1+2N_2+3N_3+4N_4+5N_5+6N_6,\notag\\
X_{3,2} &=W_{3,2} + N_1+3N_2+4N_3+5N_4+6N_5+7N_6,\notag\\
X_{3,3} &=W_{3,3} -(3N_1+6N_2+8N_3+10N_4+12N_5+14N_6).
\end{align}

Each server $ u$ broadcasts $Y_u = \sum_{v=1}^3 X_{u,v}$ to other servers. \Ip,
\begin{align}
\label{eq: Y1, Y2, example 2}
Y_1 &=W_{1,1} +W_{1,2}+W_{1,3}+ N_1+N_2+N_3,\notag\\
Y_2 &=W_{2,1} +W_{2,2}+W_{2,3} +N_4+N_5+N_6,\notag\\
Y_3 &=W_{3,1} +W_{3,2}+W_{3,3}-(N_1+N_2+N_3+N_4+N_5+N_6).
\end{align}
With $L_X=L_Y=L_Z=1$ and $L_{Z_{\Sigma}}=6$, the achieved communication rates are
$R_X=R_Y=R_Z=1$ and $R_{Z_{\Sigma}}=6$.
In this setup, each server $k\in[3]$ is capable of obtaining the total sum of all user inputs
by processing the messages from its associated users together with the messages exchanged from the other servers.
Concretely, server $k$ forms the sum of the local user messages
$\{X_{k,v}\}_{v\in[3]}$ from users $(k,1)$, $(k,2)$, $(k,3)$,
and the server-to-server messages $\{Y_u\}_{u\in[3]\setminus\{k\}}$ received from the other two servers.
Mathematically, this operation satisfies $\sum_{v=1}^{3} X_{k,v} + \sum_{u\in[3]\setminus\{k\}} Y_u = \sum_{(u,v)\in [3]\times [3]} W_{u,v}.$
For illustration, take server $k=1$ as an example: $X_{1,1}+X_{1,2}+X_{1,3}+Y_2+Y_3= (W_{1,1}+N_1) + (W_{1,2}+N_2) + (W_{1,3}+N_3)  + (W_{2,1}+W_{2,2}+W_{2,3}+N_4+N_5+N_6)  + (W_{3,1}+W_{3,2}+W_{3,3}-(N_1+N_2+N_3+N_4+N_5+N_6)) = \sum_{(u,v)\in [3]\times [3]} W_{u,v},$ since all key contributions cancel according to the chosen key design in (\ref{eq: indiv keys, example 2}).
The security of this construction is discussed in the following section.

An important property of the key construction in (\ref{eq: indiv keys, example 2}) is that
\emph{any $U+V+T-2=6$ out of the total $UV=9$ keys are mutually independent}.
Although each server observes $V+(U-1)=5$ masked messages, one linear combination of these
observations corresponds to the desired aggregation $\sum_{(u,v)\in [3]\times [3]} W_{u,v}$,
which is intentionally revealed by the protocol.
Consequently, conditioned on the desired aggregation $\sum_{(u,v)\in [3]\times [3]} W_{u,v}$
the remaining observations span at most $U+V-2=4$ independent linear dimensions.

Moreover, when up to $T=2$ users collude with the server, the colluding users reveal at most
$T$ additional input symbols.
Therefore, conditioned on the desired aggregation and the inputs of the colluding users,
the remaining observations span at most $U+V+T-2=6$ independent linear dimensions.
Since any $6$ keys among $\{Z_{u,v}\}_{(u,v)\in[3]\times[3]}$ are mutually independent,
these dimensions are fully protected by independent randomness.
As a result, conditioned on the sum $\sum_{(u,v)\in [3]\times [3]} W_{u,v}$
and the colluding users' inputs, the joint distribution of the observations
$\{X_{k,v}\}_{v\in [3]} \quad \text{and} \quad \{Y_u\}_{u\in[3]\setminus\{k\}}$
is statistically independent of the remaining individual inputs
$\{W_{u,v}\}_{(u,v)\in [3]\times [3]}$.
Hence, apart from the required aggregate and the colluding users' inputs, the server cannot infer
any additional information about the inputs.
We next formalize the above intuition. Consider Server~$1$ colluding with Users~$(1,1)$ and $(2,1)$.
\begin{align}
& I\left(X_{1,1},X_{1,2},X_{1,3}, Y_2,Y_3; \{W_{u,v}\}_{(u,v)\in[3]\times [3]}\Bigg|\sum_{(u,v)\in[3]\times [3]}W_{u,v},W_{1,1}, Z_{1,1},W_{2,1}, Z_{2,1}  \right)\notag\\
=&H\left(X_{1,1},X_{1,2},X_{1,3}, Y_2,Y_3\Bigg|\sum_{(u,v)\in[3]\times [3]}W_{u,v},W_{1,1}, Z_{1,1},W_{2,1}, Z_{2,1}  \right)\notag\\
&-H\left(X_{1,1},X_{1,2},X_{1,3}, Y_2,Y_3\Bigg|\sum_{(u,v)\in[3]\times [3]}W_{u,v},W_{1,1}, Z_{1,1},W_{2,1}, Z_{2,1}, \{W_{u,v}\}_{(u,v)\in[3]\times [3]}  \right)\\
\overset{(\ref{eq: X1,X2,X3, example 1})(\ref{eq: Y1, Y2, example 1})}{=}&H\left(W_{1,1}+N_1,W_{1,2}+N_2,W_{1,3}+N_3,W_{3,1} +W_{3,2}+W_{3,3}-(N_1+N_2+N_3+N_4+N_5+N_6) ,\right.\notag\\
&\left.W_{2,1} +W_{2,2} +W_{2,3} +N_4+N_5+ N_6\Big|
\sum_{(u,v)\in[3]\times [3]}W_{u,v},W_{1,1}, N_{1},W_{2,1}, N_{4}  \right)\notag\\
&-H\left(W_{1,1}+N_1,W_{1,2}+N_2,W_{1,3}+N_3,W_{3,1} +W_{3,2}+W_{3,3}-(N_1+N_2+N_3+N_4+N_5+N_6) ,\right.\notag\\
&\left.W_{2,1} +W_{2,2} +W_{2,3} +N_4+N_5+ N_6\Big|
\sum_{(u,v)\in[3]\times [3]}W_{u,v},W_{1,1}, N_{1},W_{2,1}, N_{4},\{W_{u,v}\}_{(u,v)\in[3]\times [3]}  \right)\\
=&H\left(W_{1,1}+N_1,W_{1,2}+N_2,W_{1,3}+N_3,W_{3,1} +W_{3,2}+W_{3,3}-(N_1+N_2+N_3+N_4+N_5+N_6) ,\right.\notag\\
&\left.W_{2,1} +W_{2,2} +W_{2,3} +N_4+N_5+ N_6,
\sum_{(u,v)\in[3]\times [3]}W_{u,v},W_{1,1}, N_{1},W_{2,1}, N_{4}  \right)\notag\\
&-H\left(\sum_{(u,v)\in[3]\times [3]}W_{u,v},W_{1,1}, N_{1},W_{2,1}, N_{4}\right)-H\left(N_2,N_3,-(N_2+N_3+N_5+N_6) ,N_5+ N_6\Big|\right.\notag\\
&\left.
\sum_{(u,v)\in[3]\times [3]}W_{u,v},W_{1,1}, N_{1},W_{2,1}, N_{4},\{W_{u,v}\}_{(u,v)\in[3]\times [3]}  \right)\\
\overset{(\ref{ind})(\ref{correctness})}{=}&H\left(W_{1,2}+N_2,W_{1,3}+N_3,W_{3,1} +W_{3,2}+W_{3,3}-(N_1+N_2+N_3+N_4+N_5+N_6) ,\right.\notag\\
&\left.W_{2,2} +W_{2,3} +N_5+ N_6,
W_{1,1}, N_{1},W_{2,1}, N_{4}  \right)-H\left(\sum_{(u,v)\in[3]\times [3]}W_{u,v},W_{1,1}, N_{1},W_{2,1}, N_{4}\right)\notag\\
&-H\left(N_2,N_3,N_5+ N_6 \right)\label{eq:ex2tt1}\\
=&8-5-3=0.
\end{align}
The first term in (\ref{eq:ex2tt1}) follows from the independence between the inputs and the keys, with $\sum_{(u,v)\in[3]\times[3]} W_{u,v}$ representing the aggregate of the remaining messages.
Similarly, the third term also follows from this independence, where the term $-(N_2+N_3+N_5+N_6)$ corresponds to a linear combination of the remaining keys $N_2,N_3,N_5+ N_6$.
\end{example}

Building on the intuition and principles illustrated by the previous examples, we now present a general construction that applies to arbitrary parameters $(U,V,T)$. This general scheme formalizes the design of the source and individual keys, as well as the communication protocol governing both user-to-server and server-to-server messages.
\subsection{General Achievability Proof of Theorem \ref{thm1}}
\label{subsec: comm & key gen sch, gen sch}
Let  the source key consist of $R_{Z_{\Sigma}}^*=  \min\{U+V+T-2,UV-1  \}$
i.i.d. uniform random variables from $\mathbb{F}_{q}$, i.e., $Z_{\Sigma}=(N_1,\cdots, N_{R_{Z_{\Sigma}}^*})$. To satisfy the field size requirements for security, we set $q > R_{Z_{\Sigma}}^* \binom{UV}{R_{Z_{\Sigma}}^*}$.
Each \indiv  key is written as a linear combination of the source key variables, i.e., 
\be
\label{eq: indiv key lcb}
Z_{u,v} =\hv_{u,v}Z_{\Sigma}^T,\quad  (u,v)\in[V]\times [U]
\ee 
where $\hv_{u,v}\in \mathbb{F}_q^{1\times \rzsigmastar   }$ is the coefficient vector, which is chosen\footnote{Our proof relies on randomized code construction and explicit codes that work universally for all cases are an interesting open problem.} as follows
\begin{eqnarray}
&& \mbox{each element of}~\{{\bf h}_{u,v}\}_{(u,v)\in ([U]\times [V])\setminus \{(U,V)\}}~\mbox{is chosen uniformly and i.i.d.} \notag\\
&& \mbox{from}~\mathbb{F}_{q}, {\bf h}_{U,V} \eqdef -\sum_{(u,v)\in ([U]\times [V])\setminus \{(U,V)\}} {\bf h}_{u,v}. \label{eq:c12}
 \end{eqnarray}
There must exist a realization\footnote{Since $R_{Z_{\Sigma}}^* < UV$, the determinant associated with any selection of $R_{Z_{\Sigma}}^*$ distinct vectors ${\bf h}_{u,v}$ defines a non-zero polynomial. Applying the Schwartz--Zippel lemma, the product of all such determinant polynomials has degree $R_{Z_{\Sigma}}^* \binom{UV}{R_{Z_{\Sigma}}^*}$. By choosing a field $\mathbb{F}_q$ with $q$ larger than this degree, the probability that the product polynomial evaluates to a non-zero value is strictly positive. This ensures the validity of (\ref{eq:sz1112222}).
} of ${\bf h}_{u,v}, (u,v) \in [U]\times [V]$ such that
\begin{eqnarray}
&&\mbox{any $R_{Z_{\Sigma}}^*$ or fewer distinct ${\bf h}_{u,v}, (u,v) \in [U]\times [V]$ and $\sum_{v\in[V]}{\bf h}_{u,v},u\in [U]$} \notag\\
&&\mbox{vectors are linearly independent.}  \label{eq:sz1112222}
\end{eqnarray}
 Moreover,
 \begin{eqnarray}
 \sum_{(u,v) \in [U]\times [V]} Z_{u,v} \overset{(\ref{eq:c12})}{=}  0. \label{eq:c131}
 \end{eqnarray}
 
User $(u,v)$ transmits the following message $X_{u,v}$ to server $u$:
\begin{equation}
\label{eq:message_Xuv_ach_scheme}
X_{u,v} = W_{u,v} + Z_{u,v},
\end{equation}
where $W_{u,v}$ is the user's input and $Z_{u,v}$ is the corresponding individual key.

Server $u$ then sums up the messages $Y_u$ received from its associated users and broadcasts to all other servers:
\begin{equation}
\label{eq:message_Yu_ach_scheme}
Y_u = \sum_{v \in [V]} X_{u,v}, \quad u \in [U].
\end{equation}

The achieved communication rates are $R_X = L_X / L = 1$ and $R_Y = L_Y / L = 1$, the individual key rate is $R_Z = L_Z / L = 1$, and the source key rate is $R_{Z_\Sigma} = L_{Z_\Sigma} / L = \min\{U+V+T-2,\, UV-1\}$, as desired.
To establish correctness, server $k$ collects the messages from its associated users, $\{X_{k,v}\}_{v\in [V]}$, as well as the messages broadcast by the other servers, $\{Y_u\}_{u\in [U]\setminus\{k\}}$, to recover the global sum of inputs:
\begin{align}
\sum_{v\in [V]} X_{k,v} + \sum_{u\in [U]\setminus\{k\}} Y_u
&= \sum_{v\in [V]} (W_{k,v} + Z_{k,v}) + \sum_{u\in [U]\setminus\{k\}} \sum_{v\in [V]} X_{u,v} \\
&= \sum_{v\in [V]} (W_{k,v} + Z_{k,v}) + \sum_{u\in [U]\setminus\{k\}} \sum_{v\in [V]} (W_{u,v} + Z_{u,v}) \\
&= \sum_{u\in [U]} \sum_{v\in [V]} (W_{u,v} + Z_{u,v}) \\
&= \sum_{(u,v)\in [U]\times [V]} W_{u,v} + \sum_{(u,v)\in [U]\times [V]} Z_{u,v} \\
&\overset{(\ref{eq:c131})}{=} \sum_{(u,v)\in [U]\times [V]} W_{u,v}.
\end{align}
Hence, the correctness is guaranteed.

Having established correctness, we now rigorously prove the security of the scheme against any set of colluding users $\mathcal{T}$ with $|\mathcal{T}|\le T$. 
The colluding user set $\mathcal{T}$ can be partitioned into three disjoint subsets:
We partition the set of colluding users $\mathcal{T}$ into three subsets for clarity and subsequent analysis:
\begin{enumerate}
    \item \textbf{Users colluding with server $k$:} 
    \begin{align}
        \mathcal{T}_1 \triangleq \mathcal{T} \cap \{(k,v)\}_{v \in [V]},\label{eqt1}
    \end{align}
    i.e., the colluding users that are associated with the specific server $k$.

    \item \textbf{Users belonging to servers for which all associated users collude:} 
    \begin{align}
        \mathcal{U}_2 \triangleq \bigl\{ u : \{(u,v)\}_{v\in [V]} \subseteq \mathcal{T},u \in [U] \bigr\}, \quad 
    \mathcal{T}_2 \triangleq \bigcup_{u \in \mathcal{U}_2} \{(u,v)\}_{v \in [V]}.\label{eqt2}
    \end{align}
    This subset contains all users from servers for which all associated users are included in the colluding set.

    \item \textbf{Remaining colluding users:} 
    \begin{align}
        \mathcal{T}_3 \triangleq \mathcal{T} \setminus (\mathcal{T}_1 \cup \mathcal{T}_2),\label{eqt3}
    \end{align}
    i.e., all colluding users not included in $\mathcal{T}_1$ or $\mathcal{T}_2$.
\end{enumerate}

We analyze the mutual information between the messages available to the server $k\in [U]$ and the global inputs, conditioned on the sum of all inputs and the colluding users' own inputs and keys:
\begin{align}
    &I\left(\{Y_u\}_{u\in[U]\setminus \{k\}},\{X_{k,v}\}_{v\in [V]}; W_{[U]\times [V]}
\Bigg|\sum_{(u,v)\in[V]\times [U]}W_{u,v}, \{W_{u,v}, Z_{u,v } \}_{(u,v )\in \Tc} \right)\notag\\
=&H\left(\{Y_u\}_{u\in[U]\setminus \{k\}},\{X_{k,v}\}_{v\in [V]}
\Bigg|\sum_{(u,v)\in[V]\times [U]}W_{u,v}, \{W_{u,v}, Z_{u,v } \}_{(u,v )\in \Tc} \right)\notag\\
&-H\left(\{Y_u\}_{u\in[U]\setminus \{k\}},\{X_{k,v}\}_{v\in [V]}
\Bigg| W_{[U]\times [V]},\sum_{(u,v)\in[V]\times [U]}W_{u,v}, \{W_{u,v}, Z_{u,v } \}_{(u,v )\in \Tc} \right)\\
 \overset{\substack{(\ref{eq:message_Xuv_ach_scheme})\\(\ref{eq:message_Yu_ach_scheme})}}{=}&H\left(\Bigg\{\sum_{v\in[V]}(W_{u,v} + Z_{u,v})\Bigg\}_{u\in[U]\setminus \{k\}},\{W_{k,v} + Z_{k,v}\}_{v\in [V]}
\Bigg|\sum_{(u,v)\in[V]\times [U]}W_{u,v}, \{W_{u,v}, Z_{u,v } \}_{(u,v )\in \Tc} \right)-\notag\\
&H\left(\Bigg\{\sum_{v\in[V]}(W_{u,v} + Z_{u,v})\Bigg\}_{u\in[U]\setminus \{k\}},\{W_{k,v} + Z_{k,v}\}_{v\in [V]}
\Bigg| W_{[U]\times [V]},\sum_{(u,v)\in[V]\times [U]}W_{u,v}, \{W_{u,v}, Z_{u,v } \}_{(u,v )\in \Tc} \right)\\
=&H\left(\Bigg\{\sum_{v\in[V]}(W_{u,v} + Z_{u,v})\Bigg\}_{u\in[U]\setminus \{k\}},\{W_{k,v} + Z_{k,v}\}_{v\in [V]},\sum_{(u,v)\in[V]\times [U]}W_{u,v}, \{W_{u,v}, Z_{u,v } \}_{(u,v )\in \Tc} \right)\notag\\
&-H\left(\sum_{(u,v)\in[V]\times [U]}W_{u,v}, \{W_{u,v}, Z_{u,v } \}_{(u,v )\in \Tc} \right)\notag\\
&-H\left(\Bigg\{\sum_{v\in[V]}Z_{u,v}\Bigg\}_{u\in[U]\setminus \{k\}},\{ Z_{k,v}\}_{v\in [V]}
\Bigg| W_{[U]\times [V]},\sum_{(u,v)\in[V]\times [U]}W_{u,v}, \{W_{u,v}, Z_{u,v } \}_{(u,v )\in \Tc} \right)\label{generaltt0}\\
\overset{\substack{(\ref{eqt1})\\(\ref{eqt2})\\(\ref{eqt3})}}{=}&H\left(\Bigg\{\sum_{v\in[V]}(W_{u,v} + Z_{u,v})\Bigg\}_{u\in[U]\setminus (\{k\}\cup \mathcal{U}_2)},\{W_{k,v} + Z_{k,v}\}_{v\in [V]}\setminus \{W_{u,v} + Z_{u,v}\}_{(u,v)\in\mathcal{T}_1}, \right.\notag\\
&\left. \{W_{u,v},Z_{u,v } \}_{(u,v )\in \Tc} \right)-H\left(\sum_{(u,v)\in[V]\times [U]}W_{u,v}, \{W_{u,v}, Z_{u,v } \}_{(u,v )\in \Tc} \right)\notag\\
&-H\left(\Bigg\{\sum_{v\in[V]}Z_{u,v}\Bigg\}_{u\in[U]\setminus (\{k\}\cup\mathcal{U}_2)},\{ Z_{k,v}\}_{v\in [V]}\setminus \{ Z_{u,v}\}_{(u,v)\in\mathcal{T}_1}
\Bigg|  \{Z_{u,v } \}_{(u,v )\in \Tc} \right)\label{generaltt1}\\
=&\Bigl((U - (1 + |\mathcal{U}_2|)) + (V - |\mathcal{T}_1|) + 2|\mathcal{T}|\Bigr) 
- \bigl(1 + 2|\mathcal{T}|\bigr) \notag\\
&-H\left(\Bigg\{\sum_{v\in[V]}Z_{u,v}\Bigg\}_{u\in[U]\setminus (\{k\}\cup\mathcal{U}_2)},\{ Z_{k,v}\}_{v\in [V]}\setminus \{ Z_{u,v}\}_{(u,v)\in\mathcal{T}_1},  \{Z_{u,v } \}_{(u,v )\in \Tc} \right)+H\left(\{Z_{u,v } \}_{(u,v )\in \Tc} \right)\\
=&\Bigl((U - (1 + |\mathcal{U}_2|)) + (V - |\mathcal{T}_1|) + 2|\mathcal{T}|\Bigr) 
- \bigl(1 + 2|\mathcal{T}|\bigr) - \Bigl((U - (1 + |\mathcal{U}_2|)) + (V - |\mathcal{T}_1|) - 1+|\mathcal{T}|\Bigr)+|\mathcal{T}|
\label{generaltt2}\\
=&0,
\end{align}
where the third term of (\ref{generaltt0}) follows from the fact that all user inputs $W_{[U]\times [V]}$ are known in the conditioning. Consequently, the conditional entropy of $\{W_{u,v}+Z_{u,v}\}$ reduces to that of the corresponding noise terms $\{Z_{u,v}\}$.

The first term in (\ref{generaltt1}) follows from the following observations.

First, the global sum $\sum_{(u,v)\in[U]\times[V]} W_{u,v}$ is fully determined by $\{\sum_{v\in[V]}(W_{u,v}+Z_{u,v})\}_{u\in[U]\setminus\{k\}},\quad
\{W_{k,v}+Z_{k,v}\}_{v\in[V]}$. Therefore, the term \(\sum_{(u,v)\in[U]\times[V]} W_{u,v}\) does not contribute additional uncertainty once these messages are given and can be removed from the entropy expression without affecting its value.

seconed, some components in the argument of the entropy are already contained in 
$\{W_{u,v}, Z_{u,v}\}_{(u,v)\in\mathcal{T}}$ and hence can be removed without affecting the joint entropy.
Specifically, by (\ref{eqt1}), the message subset 
$\{W_{u,v}+Z_{u,v}\}_{(u,v)\in \mathcal{T}_1} \subseteq \{W_{k,v}+Z_{k,v}\}_{v\in[V]}$
is fully determined by 
$\{W_{u,v}, Z_{u,v}\}_{(u,v)\in\mathcal{T}}$ because $\mathcal{T}_1\subseteq \mathcal{T}$.
Therefore, the information carried by $\{W_{u,v}+Z_{u,v}\}_{(u,v)\in \mathcal{T}_1}$ can be excluded, yielding 
$\{W_{k,v}+Z_{k,v}\}_{v\in[V]}\setminus \{W_{u,v}+Z_{u,v}\}_{(u,v)\in \mathcal{T}_1}$.

Third, for any server $u\in\mathcal{U}_2$, all users associated with server $u$ belong to the colluding set $\mathcal{T}_2$.
In this case, by (\ref{eqt2}), the aggregated message $Y_u=\sum_{v\in[V]}(W_{u,v}+Z_{u,v})$
is completely determined by 
$\{W_{u,v}, Z_{u,v}\}_{(u,v)\in\mathcal{T}}$.
Hence, the messages indexed by $\mathcal{U}_2$ in 
$\{\sum_{v\in[V]}(W_{u,v}+Z_{u,v})\}_{u\in[U]\setminus\{k\}}$
can be removed, leaving only those indexed by 
$[U]\setminus(\{k\}\cup\mathcal{U}_2)$. 

Finally, consider the subset $\mathcal{T}_3$. For each server $u$, only a subset of the users associated with server $u$ belongs to $\mathcal{T}_3$.
According to (\ref{eqt3}), the aggregated message $Y_u=\sum_{v\in[V]}(W_{u,v}+Z_{u,v})$
depends not only on the inputs of users in $\mathcal{T}_3$ but also on those of users outside $\mathcal{T}_3$.
Consequently, $Y_u$ is not fully determined by $\{W_{u,v}, Z_{u,v}\}_{(u,v)\in\mathcal{T}}$.
Therefore, the information corresponding to $Y_u=\sum_{v\in[V]}(W_{u,v}+Z_{u,v})$
cannot be removed from the entropy expression and must be retained.

The third term of (\ref{generaltt1}) follows by the same reasoning as for the previous entropy simplification. Specifically, once the contributions of the colluding users are accounted for, certain terms in the conditional entropy-namely, the messages 
\(\{Z_{u,v}\}_{(u,v)\in \mathcal{T}_1}\) and the aggregated sums 
\(\{\sum_{v\in[V]} Z_{u,v}\}_{u\in\mathcal{U}_2}\) are fully determined by 
\(\{Z_{u,v}\}_{(u,v)\in \mathcal{T}}\) and thus carry no additional uncertainty. Consequently, these terms can be removed, leaving only the messages that are not already determined by the colluding users' keys.

The third term in \eqref{generaltt2} involves a subtraction of one due to the zero-sum constraint in \eqref{eq:c131}. Even after removing the keys corresponding to the colluding users, this zero-sum property still holds. As a result, one of the remaining keys can be expressed as a linear combination of the others, which reduces the number of independent keys by one. This justifies the subtraction in the term $(U - (1 + |\mathcal{U}_2|)) + (V - |\mathcal{T}_1|) - 1+|\mathcal{T}|.$ Moreover, this term can be rewritten as $(U - (1 + |\mathcal{U}_2|)) + (V - |\mathcal{T}_1|) - 1 + |\mathcal{T}| = U + V + |\mathcal{T}| - 2 - (|\mathcal{U}_2| + |\mathcal{T}_1|) \le R_{Z_\Sigma}^*.$
According to \eqref{eq:sz1112222}, these keys are independent, which justifies the above count of independent keys.

This calculation shows that, conditioned on the sum of all inputs and the colluding users' own information, the messages observed by the colluding set $\mathcal{T}$ reveal no additional information about other users' inputs. Therefore, the scheme satisfies the security requirement.

\section{Converse Proof of Theorem \ref{thm1}}
\label{sec: converse}
In this section, we establish lower bounds on the communication rates $R_X$ and $R_Y$, as well as the key rates $R_Z$ and $R_{Z_{\Sigma}}$, by employing information-theoretic arguments. Since these bounds coincide with the achievable rates presented in Section~\ref{sec:genenralscheme}, the optimality of the proposed scheme is thereby confirmed.

Before proceeding to the general converse proof, we first consider Example~\ref{example: U=3, V=3, T=2} to illustrate the key ideas.
\subsection{Converse Proof of Example~\ref{example: U=3, V=3, T=2}}
Consider the setting in Example~\ref{example: U=3, V=3, T=2}. We show that the communication rates satisfy $R_X \ge 1$ and $R_Y \ge 1$, the individual key rate satisfies $R_Z \ge 1$, and the source key rate satisfies $R_{Z_\Sigma} \ge U+V+T-2 = 6$. For any server $k$, there are $V$ associated users, each employing an independent individual key to satisfy the security constraint, while the remaining $U-1$ servers also contribute $U-1$ independent keys.
In addition, up to $T$ colluding users may contribute $T$ independent keys. Consequently, a total of $V+(U-1)+T$ keys are involved. However, since the global aggregation is intentionally revealed to each server, these keys cannot be fully independent. In particular, one linear dependence is necessary to recover the aggregate, which effectively reduces the number of independent keys by one. Therefore, the source key rate is lower bounded by $U+V+T-2$.

We first present an auxiliary result: the entropy of $X_{1,1}$, conditioned on the other users' inputs and keys, is at least $L$ symbols:
\begin{eqnarray}
\label{example2X11}
H(X_{1,1}) \ge H\left(X_{1,1} \mid \{W_{u,v},Z_{u,v}\}_{(u,v)\in [3]\times[3]\setminus\{(1,1)\}}\right) \ge L.
\end{eqnarray}
The underlying reasoning is as follows. Every server $k \in [3]$ must be able to reconstruct the sum of all users' inputs, which contains $W_{1,1}$ in particular. Since $W_{1,1}$ is held solely by User $(1,1)$, any information about this symbol must be conveyed via the transmitted message $X_{1,1}$. Therefore, the entropy of $X_{1,1}$ is lower bounded by the entropy of $W_{1,1}$, i.e., $L$. Formally, this argument can be expressed as
\begin{eqnarray}
&& H\left( X_{1,1}|\{W_{u,v},Z_{u,v} \}_{(u,v)\in [3]\times [3]\backslash \{(1,1)\}}  \right)\notag\\
 \geq&& I\left(X_{1,1};  \sum_{(u,v)\in [3]\times [3]}W_{u,v}     \bigg |    \{W_{u,v},Z_{u,v} \}_{(u,v)\in [3]\times [3]\backslash \{(1,1)\}}          \right)\\ 
  =&& H\left(\sum_{(u,v)\in [3]\times [3]}W_{u,v}  \bigg |    \{W_{u,v},Z_{u,v} \}_{(u,v)\in [3]\times [3]\backslash \{(1,1)\}}   \right)\notag\\
 &&  - H\left(\sum_{(u,v)\in [3]\times [3]}W_{u,v}  \bigg | X_{1,1},  \{W_{u,v},Z_{u,v} \}_{(u,v)\in [3]\times [3]\backslash \{(1,1)\}}   \right)\\
\overset{(\ref{ind})(\ref{messageX})(\ref{messageY})}{=}&&    H\left(W_{1,1}   \right) - \underbrace{ H\left(\sum_{(u,v)\in [3]\times [3]}W_{u,v}  \bigg | X_{1,1},X_{1,2},X_{1,3},  \{W_{u,v},Z_{u,v} \}_{(u,v)\in [3]\times [3]\backslash \{(1,1)\}} , Y_{2}, Y_{3} \right)}_{ \overset{(\ref{correctness})}{=}    0}\label{pf_ex2_t1}\\
    \overset{(\ref{h2})}{=}&& L,\label{pf_ex2_t2}
\end{eqnarray}
where the first term in (\ref{pf_ex2_t1}) follows from the independence of the input $W_{1,1}$ and the remaining inputs and keys 
$\{W_{u,v},Z_{u,v}\}_{(u,v)\in[3]\times[3]\backslash\{(1,1)\}}$, as stated in (\ref{ind}). 
The second term in (\ref{pf_ex2_t1}) follows since 
$\{X_{u,v}\}_{(u,v)\in[3]\times[3]\backslash\{(1,1)\}}$ is fully determined by 
$\{W_{u,v},Z_{u,v}\}_{(u,v)\in[3]\times[3]\backslash\{(1,1)\}}$ according to (\ref{messageX}), 
and the server messages $Y_2$ and $Y_3$ are functions of 
$\{X_{2,v}\}_{v\in[3]}$ and $\{X_{3,v}\}_{v\in[3]}$, respectively, by (\ref{messageY}). 
Finally, in (\ref{pf_ex2_t2}), we use the fact that $W_{1,1}$ consists of $L$ independent and uniformly distributed symbols (see (\ref{h2})). 
Moreover, by the correctness condition, the desired sum 
$\sum_{(u,v)\in[3]\times[3]} W_{u,v}$ can be decoded with no error from the messages 
$X_{1,1},X_{1,2},X_{1,3},Y_{2},Y_{3}$ (see (\ref{correctness})).

Next, we show that the entropy of $Y_{1}$, conditioned on the inputs and keys of all users except User~$(1,1)$ associated with server~1, is no less than $L$ symbols.
\begin{align}
H(Y_1)\geq H\left( Y_1 ,\middle|, \{W_{u,v},Z_{u,v}\}_{(u,v)\in ([3]\times [3])\setminus \{(1,1)\}} \right)
\ge L.
\label{example2Y1}
\end{align}
The underlying reasoning is as follows. Every server $k\in [3]\setminus\{1\}$ must be able to reconstruct the sum of all users' inputs, which necessarily includes the symbol $W_{1,1}$. Since $W_{1,1}$ is known exclusively to User~$(1,1)$, any information about this symbol must be conveyed through the message $Y_1$ once the inputs and keys of all other users are given. Consequently, the entropy of $Y_1$, conditioned on $\{W_{u,v}, Z_{u,v}\}_{(u,v)\in ([3]\times[3])\setminus\{(1,1)\}}$, cannot be smaller than the entropy of $W_{1,1}$, namely $L$.
The formal argument is given below.
\begin{eqnarray}
&& H\left( Y_1|\{W_{u,v},Z_{u,v} \}_{(u,v)\in [3]\times [3]\backslash \{(1,1)\}}  \right)\notag\\
 \geq&& I\left( Y_1 ; \sum_{(u,v)\in [3]\times [3]}W_{u,v}    \bigg|\{W_{u,v},Z_{u,v}\}_{(u,v)\in [3]\times [3]\backslash \{(1,1)\}}  \right)\\
 =&& H\left( \sum_{(u,v)\in [3]\times [3]}W_{u,v}    \bigg|\{W_{u,v},Z_{u,v}\}_{(u,v)\in [3]\times [3]\backslash \{(1,1)\}}  \right)\notag\\
 &&  
 - H\left( \sum_{(u,v)\in [3]\times [3]}W_{u,v}    \bigg|Y_1,\{W_{u,v},Z_{u,v}\}_{(u,v)\in [3]\times [3]\backslash \{(1,1)\}}  \right)\\
 =&& H(W_{1,1})- \underbrace{H\left( \sum_{(u,v)\in [3]\times [3]}W_{u,v}    \bigg|Y_1,Y_2,X_{3,1},X_{3,2},X_{3,3},\{W_{u,v},Z_{u,v}\}_{(u,v)\in [3]\times [3]\backslash \{(1,1)\}}  \right)}_{\overset{(\ref{messageX})(\ref{messageY}) (\ref{correctness})}{=}0}\label{pf_ex2_t222}\\
 =&L,
\end{eqnarray}
where the second term in (\ref{pf_ex2_t222}) follows from the following facts. 
First, the messages 
$\{X_{u,v}\}_{(u,v)\in[3]\times[3]\backslash\{(1,1)\}}$ are completely determined by 
$\{W_{u,v},Z_{u,v}\}_{(u,v)\in[3]\times[3]\backslash\{(1,1)\}}$ according to (\ref{messageX}). 
Second, the server message $Y_2$ is a function of $\{X_{2,v}\}_{v\in[3]}$, as specified in (\ref{messageY}). 
Finally, by the correctness condition (\ref{correctness}), the desired sum 
$\sum_{(u,v)\in[3]\times[3]} W_{u,v}$ can be recovered from 
$Y_1,Y_2,X_{3,1},X_{3,2},X_{3,3}$.

With the above preparations, we can now establish the lower bounds on the communication rates. 
From the previous entropy calculations, we have shown that each input symbol $W_{1,1}$ carries $L$ uniform bits and that the corresponding messages $X_{1,1}$ and $Y_1$ preserve this information under the encoding and aggregation process. 
Consequently, the entropies of $X_{1,1}$ and $Y_1$ provide immediate lower bounds on the required communication. 
Specifically, we have
\begin{align}
    R_X=\frac{L_X}{L} =\frac{H(X_{1,1})}{L}\overset{(\ref{example2X11})}{\geq} 1, \quad
    R_Y=\frac{L_Y}{L} =\frac{H(Y_{1})}{L}\overset{(\ref{example2Y1})}{\geq} 1,
\end{align}
which follow directly from the fact that each message must at least convey the entropy of the corresponding input to satisfy correctness and security.

Next, we show that the entropy of the messages associated with Server~$1$, namely $X_{1,1}, X_{1,2}, X_{1,3}$, is at least $UL = 3L$, even when conditioned on the inputs and keys of the colluding users. Without loss of generality, we assume that the colluding users are $(2,1)$ and $(3,1)$. Formally, we have
\begin{align}
H\left(X_{1,1},X_{1,2},X_{1,3}|\{W_{u,v},Z_{u,v}\}_{(u,v)\in \Tc}\right)\ge 3L. \label{excovX}
\end{align}
The proof proceeds as follows.
 \begin{align}
        &H\left(X_{1,1},X_{1,2},X_{1,3}|\{W_{u,v},Z_{u,v}\}_{(u,v)\in \Tc}\right)\notag\\
        =&H\left(X_{1,1},X_{1,2},X_{1,3}|W_{2,1},W_{3,1},Z_{2,1},Z_{3,1}\right)\\   =&H\left(X_{1,1}|W_{2,1},W_{3,1},Z_{2,1},Z_{3,1}\right)+H\left(X_{1,2}|X_{1,1},W_{2,1},W_{3,1},Z_{2,1},Z_{3,1}\right)\notag\\
        &+H\left(X_{1,3}|X_{1,1},X_{1,2},W_{2,1},W_{3,1},Z_{2,1},Z_{3,1}\right)\\    \geq&H\left(X_{1,1}|W_{2,1},W_{3,1},Z_{2,1},Z_{3,1}\right)+H\left(X_{1,2}|W_{1,1},Z_{1,1},X_{1,1},W_{2,1},W_{3,1},Z_{2,1},Z_{3,1}\right)\notag\\
        &+H\left(X_{1,3}|W_{1,1},Z_{1,1},X_{1,1},W_{1,2},Z_{1,2},X_{1,2},W_{2,1},W_{3,1},Z_{2,1},Z_{3,1}\right)\\        
        \overset{(\ref{messageX})}{=}&H\left(X_{1,1}|W_{2,1},W_{3,1},Z_{2,1},Z_{3,1}\right)+H\left(X_{1,2}|W_{1,1},Z_{1,1},W_{2,1},W_{3,1},Z_{2,1},Z_{3,1}\right)\notag\\
        &+H\left(X_{1,3}|W_{1,1},Z_{1,1},W_{1,2},Z_{1,2},W_{2,1},W_{3,1},Z_{2,1},Z_{3,1}\right)\\       
        \geq&H\left(X_{1,1}|\{W_{u,v},Z_{u,v}\}_{(u,v)\in([3]\times [3])\setminus\{(1,1)\}}\right)+H\left(X_{1,2}|\{W_{u,v},Z_{u,v}\}_{(u,v)\in([3]\times [3])\setminus\{(1,2)\}}\right)\notag\\
        &+H\left(X_{1,3}|\{W_{u,v},Z_{u,v}\}_{(u,v)\in([3]\times [3])\setminus\{(1,3)\}}\right)\\
        \overset{(\ref{example2X11})}{\geq}&3L.
    \end{align}

Having established the entropy lower bound for the messages associated with Server~$1$, we now consider the messages from the remaining servers, $Y_2$ and $Y_3$. We show that their joint entropy is at least $(V-1)L = 2L$, even when conditioned on the messages $X_{1,1}, X_{1,2}, X_{1,3}$ sent by the users associated with Server~$1$, as well as the inputs and keys of the colluding users. Without loss of generality, we assume that the colluding users are $(2,1)$ and $(3,1)$.
\begin{align}
H\Big(Y_2,Y_3|X_{1,1},X_{1,2},X_{1,3},\{W_{u,v},Z_{u,v}\}_{(u,v)\in \Tc}\Big)\ge 2L. \label{excovY}
\end{align}
The proof proceeds as follows.
\begin{align}
    &H\left(Y_2,Y_3 |X_{1,1},X_{1,2},X_{1,3},\{W_{u,v},Z_{u,v}\}_{(u,v)\in \Tc}\right)\notag\\
    =&H\left(Y_2,Y_3 |X_{1,1},X_{1,2},X_{1,3},W_{2,1},W_{3,1},Z_{2,1},Z_{3,1}\right)\\
    \geq&H\left(Y_2 |X_{1,1},X_{1,2},X_{1,3},W_{2,1},W_{3,1},Z_{2,1},Z_{3,1}\right)\notag\\
    &+H\left(Y_3 |Y_2,X_{1,1},X_{1,2},X_{1,3},W_{2,1},W_{3,1},Z_{2,1},Z_{3,1}\right)\\
    \geq&H\left(Y_2 |\{W_{1,v},Z_{1,v}\}_{v\in [3]},X_{1,1},X_{1,2},X_{1,3},W_{2,1},W_{3,1},Z_{2,1},Z_{3,1}\right)\notag\\
    &+H(Y_3 |\{W_{2,v},Z_{2,v}\}_{v\in [3]},Y_2,\{W_{1,v},Z_{1,v}\}_{v\in [3]},X_{1,1},X_{1,2},X_{1,3},W_{2,1},W_{3,1},Z_{2,1},Z_{3,1})\\
    =&H\left(Y_2 |\{W_{1,v},Z_{1,v}\}_{v\in [3]},W_{2,1},W_{3,1},Z_{2,1},Z_{3,1}\right)\notag\\
    &+H(Y_3 |\{W_{2,v},Z_{2,v}\}_{v\in [3]},\{W_{1,v},Z_{1,v}\}_{v\in [3]},W_{2,1},W_{3,1},Z_{2,1},Z_{3,1})\\
    \geq&H\left(Y_2 |\{W_{u,v},Z_{u,v}\}_{(u,v)\in ([3]\times[3])\setminus\{(2,2)\}}\right)+H(Y_3 |\{W_{u,v},Z_{u,v}\}_{(u,v)\in ([3]\times[3])\setminus\{(3,2)\}})\\
    \overset{(\ref{example2Y1})}{\geq}&2L.
\end{align}

Building on the previous entropy lower bounds for the messages, we now consider the security requirements imposed by colluding users. In particular, to maintain security, the keys held by colluding users must be independent. Specifically, we have
\begin{align}
 H(Z_{2,1}, Z_{3,1}) \ge 2L. \label{excovkey}
\end{align}

The proof proceeds as follows. Using the chain rule of entropy, $H(Z_{2,1}, Z_{3,1}) = H(Z_{2,1}) + H(Z_{3,1} \mid Z_{2,1}),$
we first show that \(H(Z_{3,1} \mid Z_{2,1}) \ge L\); the bound for \(H(Z_{2,1})\) can be shown similarly.
\begin{align}
H(Z_{3,1}) 
&\ge H(Z_{3,1} \mid Z_{2,1}) \label{excot1} \\
&\ge H(Z_{3,1} \mid W_{3,1}, Z_{2,1}, W_{2,1}) \\
&\ge H(Z_{3,1}; X_{3,1} \mid W_{3,1}, Z_{2,1}, W_{2,1}) \\
&\overset{(\ref{messageX})}{=} H(X_{3,1} \mid W_{3,1}, Z_{2,1}, W_{2,1}) \\
&= H(X_{3,1} \mid Z_{2,1}, W_{2,1}) - I(X_{3,1}; W_{3,1} \mid Z_{2,1}, W_{2,1}) \\
&\ge H(X_{3,1} \mid Z_{2,1}, W_{2,1}) - I\Bigg(X_{3,1}, \sum_{(u,v)\in[3]\times[3]} W_{u,v}; W_{3,1} \Bigg| Z_{2,1}, W_{2,1}\Bigg) \\
&\overset{(\ref{example2X11})}{\ge} L 
 - \underbrace{I\Bigg(\sum_{(u,v)\in[3]\times[3]} W_{u,v}; W_{3,1} \Bigg| Z_{2,1}, W_{2,1}\Bigg)}_{\overset{(\ref{ind})}{=} 0} 
 - I\Bigg(X_{3,1}; W_{3,1} \Bigg| \sum_{(u,v)\in[3]\times[3]} W_{u,v}, Z_{2,1}, W_{2,1}\Bigg) \label{ex2covtt1} \\
&\ge L - \underbrace{I\Bigg(X_{3,1}, X_{3,2}, X_{3,3}, Y_1, Y_2; \{W_{u,v}\}_{(u,v)\in[3]\times[3]} \Bigg| \sum_{(u,v)\in[3]\times[3]} W_{u,v}, Z_{2,1}, W_{2,1}\Bigg)}_{\overset{(\ref{security})}{=} 0} \\
&= L, \label{excot2}
\end{align}

where the second terms in (\ref{ex2covtt1}) vanish due to the mutual independence of inputs and keys.

Similarly, we can show that
\begin{align}
H(Z_{2,1}) \ge L.
\end{align}

Having established the independence of the colluding users' keys and the corresponding entropy lower bounds, we are now ready to quantify the individual key rate. In particular, we show that
\begin{align}
R_Z = \frac{L_Z}{L} = \frac{H(X_{3,1})}{L} \overset{(\ref{excot2})}{\geq} 1.
\end{align}

Having established the bounds in (\ref{excovX}), (\ref{excovY}), and (\ref{excovkey}), we can now turn to the converse proof for the source key rate, demonstrating the minimal key requirement necessary to maintain security in the presence of colluding users.
\begin{align}
    &H(Z_{\Sigma})\\
    \overset{(\ref{sourcekey})}{\geq}&H(\{Z_{u,v}\}_{(u,v)\in[3]\times [3]})\\
    =&H(\{Z_{u,v}\}_{(u,v)\in \Tc})+H(\{Z_{u,v}\}_{(u,v)\in[3]\times [3]}|\{Z_{u,v}\}_{(u,v)\in \Tc})\\
    \overset{(\ref{excovkey})}{\geq}&2L+H(\{Z_{u,v}\}_{(u,v)\in[3]\times [3]}|\{W_{u,v}\}_{(u,v)\in[3]\times [3]},\{Z_{u,v}\}_{(u,v)\in \Tc})\\
    \geq&2L+I(\{Z_{u,v}\}_{(u,v)\in[3]\times [3]};X_{1,1},X_{1,2},X_{1,3},Y_2,Y_3|\{W_{u,v}\}_{(u,v)\in[3]\times [3]},\{Z_{u,v}\}_{(u,v)\in \Tc})\\
    =&2L+H(X_{1,1},X_{1,2},X_{1,3},Y_2,Y_3|\{W_{u,v}\}_{(u,v)\in[3]\times [3]},\{Z_{u,v}\}_{(u,v)\in \Tc})\notag\\
    &-\underbrace{H(X_{1,1},X_{1,2},X_{1,3},Y_2,Y_3|\{Z_{u,v}\}_{(u,v)\in[3]\times [3]},\{W_{u,v}\}_{(u,v)\in[3]\times [3]},\{Z_{u,v}\}_{(u,v)\in \Tc})}_{ \overset{(\ref{messageX}) (\ref{messageY})}{=}0 }\\
    =&2L+H(X_{1,1},X_{1,2},X_{1,3},Y_2,Y_3|\{Z_{u,v}\}_{(u,v)\in \Tc})\notag\\
    &-H(X_{1,1},X_{1,2},X_{1,3},Y_2,Y_3;\{W_{u,v}\}_{(u,v)\in[3]\times [3]}|\{Z_{u,v}\}_{(u,v)\in \Tc})\\
    \geq&2L+H(X_{1,1},X_{1,2},X_{1,3}|\{Z_{u,v}\}_{(u,v)\in \Tc})+H(Y_2,Y_3|X_{1,1},X_{1,2},X_{1,3},\{Z_{u,v}\}_{(u,v)\in \Tc})\notag\\
    &-H\left(X_{1,1},X_{1,2},X_{1,3},Y_2,Y_3,\sum_{(u,v)\in [3]\times [3]}W_{u,v};\{W_{u,v}\}_{(u,v)\in[3]\times [3]}\Bigg|\{Z_{u,v}\}_{(u,v)\in \Tc}\right)\\
    \overset{(\ref{excovX})(\ref{excovY})}{\geq}&2L+3L+2L-H\left(\sum_{(u,v)\in [3]\times [3]}W_{u,v};\{W_{u,v}\}_{(u,v)\in[3]\times [3]}\Bigg|\{Z_{u,v}\}_{(u,v)\in \Tc}\right)\notag\\
    &-\underbrace{H\left(X_{1,1},X_{1,2},X_{1,3},Y_2,Y_3;\{W_{u,v}\}_{(u,v)\in[3]\times [3]}\Bigg|\sum_{(u,v)\in [3]\times [3]}W_{u,v},\{Z_{u,v}\}_{(u,v)\in \Tc}\right)}_{ \overset{(\ref{security})}{=}0 }\label{excovsourt1}\\
    =&2L+3L+2L-L=6L,
\end{align}
The fourth term in (\ref{excovsourt1}) is $L$, as the sum conveys exactly one symbol of information about the inputs, which is independent of the keys held by the colluding users.

The preceding example illustrates the key ideas and techniques used to derive entropy lower bounds and establish key independence. We now extend these arguments to the general setting, providing the converse proof of Theorem~\ref{thm1}.

\subsection{General Converse Proof of Theorem \ref{thm1}}
\label{subsec: feasible region, converse}

We now generalize the previous proof to arbitrary parameter settings. To this end, we first introduce several useful lemmas. As a preliminary, we show that each message $X_{u,v}$ must carry at least $L$ symbols-the size of the input-even when all other inputs are known. Similarly, each relay message $Y_u$ must contain at least $L$ symbols, provided that at least one associated input $X_{u,v}$ remains unknown. In all lemmas, the colluding user set $\mathcal{T}$ satisfies $|\mathcal{T}|\le T\leq (U-1)(V-1)$.

\begin{lemma}
\label{lemma: message info lemma}
\emph{For any $(u,v) \in [U] \times [V]$, it holds that}
\begin{align}
& H\left( X_{u,v}|\{W_{i,j},Z_{i,j} \}_{(i,j)\in [U]\times [V]\backslash \{(u,v)\}}  \right)
\ge L,\label{lemma1X>=L}\\
& H\left( Y_u|\{W_{i,j},Z_{i,j} \}_{(i,j)\in [U]\times [V]\backslash \{(u,v)\}}  \right)
\ge L\label{lemma1Y>=L}.
\end{align}
\end{lemma}

\begin{IEEEproof}
A server can reconstruct the aggregate $\sum_{(u,v)\in [U]\times[V]} W_{u,v}$ only if the information of every input symbol $W_{u,v}$ is conveyed through the user-to-server and server-to-server links. Hence, the messages transmitted over these links must contain sufficient information to uniquely determine $W_{u,v}$. Since $H(W_{u,v}) = L$, the entropy of the corresponding message cannot be smaller than $L$ bits.  The formal justification follows from (\ref{lemma1X>=L}).
\begin{align}
& H\left( X_{u,v}|\{W_{i,j},Z_{i,j} \}_{(i,j)\in [U]\times [V]\backslash \{(u,v)\}}  \right)\notag\\
 \geq& I\left(X_{u,v};  \sum_{(i,j)\in [U]\times [V]}W_{i,j}     \bigg |    \{W_{i,j},Z_{i,j} \}_{(i,j)\in [U]\times [V]\backslash \{(u,v)\}}          \right)\\ 
  =& H\left(\sum_{(i,j)\in [U]\times [V]}W_{i,j}  \bigg |    \{W_{i,j},Z_{i,j} \}_{(i,j)\in [U]\times [V]\backslash \{(u,v)\}}   \right)\notag\\
 &  - H\left(\sum_{(i,j)\in [U]\times [V]}W_{i,j}  \bigg | X_{u,v},  \{W_{i,j},Z_{i,j} \}_{(i,j)\in [U]\times [V]\backslash \{(u,v)\}}   \right)\\
\overset{(\ref{ind})(\ref{messageX})(\ref{messageY})}{=}&    H\left(W_{u,v}   \right) - \underbrace{ H\left(\sum_{(i,j)\in [U]\times [V]}W_{i,j}  \bigg | \{X_{u,v}\}_{v\in [V]},  \{W_{i,j},Z_{i,j} \}_{(i,j)\in [U]\times [V]\backslash \{(u,v)\}} , \{Y_k\}_{k\in{[U]\setminus\{u\}}} \right)}_{ \overset{(\ref{correctness})}{=}    0}\label{pf_lemma1_t1}\\
    \overset{(\ref{h2})}{=}& L,\label{pf_lemma1_t2}
\end{align}
The first term in (\ref{pf_lemma1_t1}) follows from the fact that the input $W_{u,v}$ is independent of all other inputs and keys $\{W_{i,j},Z_{i,j}\}_{(i,j) \in [U]\times [V] \setminus \{(u,v)\}}$ (see (\ref{ind})). The second term in (\ref{pf_lemma1_t1}) follows from the fact that $\{X_{i,j}\}_{(i,j)\in [U]\times [V] \setminus \{(u,v)\}}$ is determined by $\{W_{i,j},Z_{i,j}\}_{(i,j)\in [U]\times [V] \setminus \{(u,v)\}}$ (see (\ref{messageX})) and $\{Y_k\}_{k\in [U]\setminus\{u\}}$ is determined by $\{X_{i,j}\}_{(i,j)\in [U]\setminus\{u\}\times [V]}$ (see (\ref{messageY})). Moreover, the second term of (\ref{pf_lemma1_t1}) is zero because the desired sum $\sum_{(u,v)\in [U]\times [V]} W_{u,v}$ can be decoded without error from the messages $\{X_{u,v}\}_{v\in [V]}$ and $\{Y_k\}_{k\in [U]\setminus\{u\}}$ (see (\ref{correctness})).  
In (\ref{pf_lemma1_t2}), we use the fact that $W_{u,v}$ consists of $L$ uniform symbols (see (\ref{h2})).

Having established the entropy lower bound for each message $X_{u,v}$ in (\ref{lemma1X>=L}), we now turn to the messages $Y_u$. The proof of (\ref{lemma1Y>=L}) follows a similar argument:
\begin{eqnarray}
&& H\left( Y_u|\{W_{i,j},Z_{i,j} \}_{(i,j)\in [U]\times [V]\backslash \{(u,v)\}}  \right)\notag\\
 \geq&& I\left( Y_u ; \sum_{(i,j)\in [U]\times [V]}W_{i,j}    \bigg|\{W_{i,j},Z_{i,j}\}_{(i,j)\in [U]\times [V]\backslash \{(u,v)\}}  \right)\\
 =&& H\left( \sum_{(i,j)\in [U]\times [V]}W_{i,j}    \bigg|\{W_{i,j},Z_{i,j}\}_{(i,j)\in [U]\times [V]\backslash \{(u,v)\}}  \right)\notag\\
 &&  
 - H\left( \sum_{(i,j)\in [U]\times [V]}W_{i,j}    \bigg|Y_u,\{W_{i,j},Z_{i,j}\}_{(i,j)\in [U]\times [V]\backslash \{(u,v)\}}  \right)\\
 =&& H(W_{u,v})- \underbrace{H\left( \sum_{(i,j)\in [U]\times [V]}W_{i,j}    \bigg|\{Y_k\}_{k\in [U]\setminus\{u'\}},\{X_{u',v'}\}_{v'\in[V]},\{W_{i,j},Z_{i,j}\}_{(i,j)\in [U]\times [V]\backslash \{(u,v)\}}  \right)}_{\overset{(\ref{messageX}) (\ref{messageY}) (\ref{correctness})}{=}0}\label{pf_lemma1_tt2}\\
 =&&L.
\end{eqnarray}
The second term in (\ref{pf_lemma1_tt2}) follows from the facts that $\{X_{i,j}\}_{(i,j)\in [U]\times [V]\setminus \{(u,v)\}}$ is completely determined by $\{W_{i,j},Z_{i,j}\}_{(i,j)\in [U]\times [V]\setminus \{(u,v)\}}$ (see (\ref{messageX})). In particular, for $u'\neq u$, $\{X_{u',v'}\}_{v'\in [V]}$ is determined by $\{W_{u',j},Z_{u',j}\}_{j\in [V]}$, which is contained in $\{W_{i,j},Z_{i,j}\}_{(i,j)\in [U]\times [V]\setminus \{(u,v)\}}$. Moreover, $\{Y_k\}_{k\in [U]\setminus\{u'\}}$ is determined by $Y_u$ together with $\{X_{i,j}\}_{(i,j)\in [U]\times [V]\setminus \{(u,v)\}}$, as specified in (\ref{messageY}).

\end{IEEEproof}

The above lemma establishes a per-message entropy lower bound. We next strengthen this result by extending it to collections of messages and conditioning on the inputs and keys of colluding users. Note that 

\begin{lemma}
\emph{
For any $u'\in[U]$, any $v'\in[V]$, any $\mathcal{U}\subseteq [U]\setminus\{u'\}$,
any $\Vc\subseteq [V]$, and any
$\Tc \subset ([U]\setminus \{u'\})\times ([V]\setminus \{v'\})$ satisfy $|\Tc|\le T\leq (U-1)(V-1)$, we have
}
\begin{align}
&H\left(\{X_{u',v}\}_{v\in\Vc}\,\middle|\,\{W_{i,j},Z_{i,j}\}_{(i,j)\in \Tc}\right)\ge |\Vc|L, \label{lemma2xvwz}\\
&H\left(\{Y_u\}_{u\in \mathcal{U}} \,\middle|\,\{X_{u',v}\}_{v\in[V]},\{W_{i,j},Z_{i,j}\}_{(i,j)\in \Tc}\right)\ge |\mathcal{U}|L. \label{lemma2yuwz}
\end{align}
\end{lemma}
\begin{IEEEproof}
Consider (\ref{lemma2xvwz}), we have
    \begin{align}
        &H\left(\{X_{u',v}\}_{v\in\Vc}\,\middle|\,\{W_{i,j}\}_{(i,j)\in \Tc},\{Z_{i,j}\}_{(i,j)\in \Tc}\right)\notag\\
        =& \sum_{v\in \Vc  }H\left( X_{u',v}| \{X_{u',l}\}_{l\in \Vc\backslash \{v\},l<v }    ,\{W_{i,j},Z_{i,j}\}_{(i,j)\in \Tc}\right)\\
 \ge& \sum_{v\in \Vc  }H\left( X_{u',v}| \{X_{u',l}\}_{l\in \Vc\backslash \{v\} }    ,\{W_{i,j},Z_{i,j}\}_{(i,j)\in \Tc}\right)\\
\ge &\sum_{v\in \Vc  }H\left( X_{u',v}|\{W_{u',l}, Z_{u',l}\}_{l\in \Vc\backslash \{v\} }, \{X_{u',l}\}_{l\in \Vc\backslash \{v\} }    ,\{W_{i,j},Z_{i,j}\}_{(i,j)\in \Tc}\right)\\
 \overset{(\ref{messageX})}{=}& \sum_{v\in \Vc  }H\left( X_{u',v}|\{W_{u',l}, Z_{u',l}\}_{l\in \Vc\backslash \{v\} }   ,\{W_{i,j},Z_{i,j}\}_{(i,j)\in \Tc}\right)\label{lemma2ttt0}\\
\geq& \sum_{v\in \Vc  }H\left( X_{u',v}|\{W_{u',l}, Z_{u',l}\}_{(u',l)\in [U]\times[V]\backslash \{(u',v)\} } \right)\label{lemma2ttt1}\\
 \overset{(\ref{lemma1X>=L})}{\ge}&|\Vc|L,
    \end{align}
where (\ref{lemma2ttt1}) follows from the fact that conditioning reduces entropy. For any $v\in\Vc$, we have $\{W_{u',l}, Z_{u',l}\}_{l\in \Vc\setminus \{v\}}$ $ \subseteq \{W_{u',l}, Z_{u',l}\}_{l\in [V]\setminus \{v\}} \subseteq \{W_{i,j}, Z_{i,j}\}_{(i,j)\in [U]\times [V]\setminus \{(u',v)\}}$, and since $(u',v)\notin \Tc$, it also holds that $\{W_{i,j},Z_{i,j}\}_{(i,j)\in \Tc}$ $ \subseteq \{W_{i,j}, Z_{i,j}\}_{(i,j)\in [U]\times [V]\setminus \{(u',v)\}}$. Therefore, the conditioning set $\{W_{u',l}, Z_{u',l}\}_{l\in \Vc\backslash \{v\} }   \cup\{W_{i,j},Z_{i,j}\}_{(i,j)\in \Tc}$ in (\ref{lemma2ttt0}) is a subset of $\{W_{i,j}, Z_{i,j}\}_{(i,j)\in [U]\times [V]\setminus \{(u',v)\}}$, which yields (\ref{lemma2ttt1}).

Consider (\ref{lemma2yuwz}), we have
\begin{align}
    &H\left(\{Y_u\}_{u\in \mathcal{U}} \,\middle|\,\{X_{u',v}\}_{v\in[V]},\{Z_{i,j}\}_{(i,j)\in \Tc}\right)\notag\\
    = & \sum_{u\in \mathcal{U}}H\left( Y_{u}| \{Y_{k}\}_{k\in \mathcal{U}\setminus \{u\},k<u } ,\{X_{u',v}\}_{v\in[V]},\{W_{i,j},Z_{i,j}\}_{(i,j)\in \Tc}\right) \\
    \geq & \sum_{u\in \mathcal{U}}H\left( Y_{u}| \{Y_{k}\}_{k\in \mathcal{U}\setminus\{u\} } ,\{X_{u',v}\}_{v\in[V]},\{W_{i,j},Z_{i,j}\}_{(i,j)\in \Tc}\right) \\
    \geq & \sum_{u\in \mathcal{U}}H\left( Y_{u}|\{W_{k,v},Z_{k,v}\}_{(k,v)\in (\mathcal{U}\setminus\{u\})\times [V]},  \{Y_{k}\}_{k\in \mathcal{U}\setminus\{u\} } ,\{W_{u',v},Z_{u',v}\}_{v\in[V]},\{X_{u',v}\}_{v\in[V]},\{W_{i,j},Z_{i,j}\}_{(i,j)\in \Tc}\right) \\
    \overset{\substack{(\ref{messageX})\\ (\ref{messageY})}}{=}& \sum_{u\in \mathcal{U}}H\left( Y_{u}|\{W_{k,v},Z_{k,v}\}_{(k,v)\in (\mathcal{U}\setminus\{u\})\times [V]} ,\{W_{u',v},Z_{u',v}\}_{v\in[V]},\{W_{i,j},Z_{i,j}\}_{(i,j)\in \Tc}\right) \label{lemma2ttt22}\\
    \geq &\sum_{u\in \mathcal{U}}H\left( Y_{u}|\{W_{k,v},Z_{k,v}\}_{(k,v)\in [U]\times [V]\setminus \{(u,v')\}} \right)\label{lemma2ttt2}\\
    \overset{(\ref{lemma1Y>=L})}{\geq}&|\mathcal{U}|L,
\end{align}
where (\ref{lemma2ttt2}) follows from the fact that conditioning reduces entropy. For any $u\in\mathcal{U}$ and $k\in \mathcal{U}\setminus\{k\}$, the variables $\{W_{k,v}, Z_{k,v}\}_{(k,v)\in (\mathcal{U}\setminus\{u\})\times [V]}$ satisfy $\{W_{k,v}, Z_{k,v}\}_{(k,v)\in (\mathcal{U}\setminus\{u\})\times [V]} \subseteq \{W_{k,v}, Z_{k,v}\}_{(k,v)\in [U]\times [V]\setminus \{(u,v')\}}$. Moreover, since $u'\neq u$, it holds that $\{W_{u',v}, Z_{u',v}\}_{v\in [V]} \subseteq \{W_{k,v}, Z_{k,v}\}_{(k,v)\in [U]\times [V]\setminus \{(u,v')\}}$. Finally, as $(u,v')\notin \Tc$, we also have $\{W_{i,j}, Z_{i,j}\}_{(i,j)\in \Tc} \subseteq \{W_{i,j}, Z_{i,j}\}_{(i,j)\in [U]\times [V]\setminus \{(u,v')\}}$. Therefore, the conditioning set in (\ref{lemma2ttt22}) is a subset of $\{W_{k,v}, Z_{k,v}\}_{(k,v)\in [U]\times [V]\setminus \{(u,v')\}}$, which yields (\ref{lemma2ttt2}).

\end{IEEEproof}

\begin{lemma}
\label{lemma3ZV>=VL}
For any $u' \in [U]$ and $v' \in [V]$, let $\Vc \subseteq [V]$ and
$\Tc \subseteq ([U]\setminus\{u'\}) \times ([V]\setminus\{v'\})$
satisfy $|\Tc| \le T \le (U-1)(V-1)$. Then the following inequalities hold:
\begin{align}
    H\Big( \{Z_{u',v}\}_{v \in \Vc} \,\big|\, \{Z_{i,j}\}_{(i,j)\in\Tc} \Big)
    &\ge |\Vc| L, \label{lemma3zv} \\
    H\Big( \{Z_{u,v}\}_{(u,v)\in\Tc} \Big)
    &\ge |\Tc| L. \label{lemma3zt}
\end{align}
\end{lemma}

\begin{IEEEproof}
To protect the inputs $\{W_{u',v}\}_{v\in \Vc}$ from Server $u'$, the \indiv keys $\{Z_{u',v}\}_{v\in \Vc}$ must be mutually \indep even if Server $u'$ knows all the \indiv keys at the colluding users. Intuitively, this is because if $\{Z_{u',v}\}_{v\in \Vc}$ are not  \indep, Server $u$ may infer certain \info about the inputs $ \{W_{u',v}\}_{v\in \Vc}  $ by exploiting the correlation among $\{Z_{u',v}\}_{v\in \Vc}$, which violates the security constraint. More specifically, consider (\ref{lemma3zv}). We have
\begin{align}
& H\left( \{Z_{u',v}\}_{v\in\Vc}|\{Z_{i,j}\}_{(i,j)\in \Tc} \right)\notag\\
 \ge & H\left( \{Z_{u',v}\}_{v\in\Vc}|\{W_{u',v}\}_{v\in\Vc},\{W_{i,j},Z_{i,j}\}_{(i,j)\in \Tc} \right)\\
 \ge & I\left( \{Z_{u',v}\}_{v\in\Vc};\{X_{u',v}\}_{v\in\Vc}   |\{W_{u',v}\}_{v\in\Vc},\{W_{i,j},Z_{i,j}\}_{(i,j)\in \Tc} \right)\\
 = &H\left(\{X_{u',v}\}_{v\in\Vc}   |\{W_{u',v}\}_{v\in\Vc},\{W_{i,j},Z_{i,j}\}_{(i,j)\in \Tc} \right)\notag\\
&   - 
\underbrace{H\left(\{X_{u',v}\}_{v\in\Vc}   |\{Z_{u',v}\}_{v\in\Vc} ,\{W_{u',v}\}_{v\in\Vc},\{W_{i,j},Z_{i,j}\}_{(i,j)\in \Tc} \right)}_{\overset{(\ref{messageX})}{=}0  }\\
 = & H\left(\{X_{u',v}\}_{v\in\Vc}   |\{W_{i,j},Z_{i,j}\}_{(i,j)\in \Tc} \right)   -I\left(\{X_{u',v}\}_{v\in\Vc};\{W_{u',v}\}_{v\in\Vc}   |\{W_{i,j},Z_{i,j}\}_{(i,j)\in \Tc} \right)\\
\overset{(\ref{lemma2xvwz})}{\geq} & |\Vc|L  -I\left(\{X_{u',v}\}_{v\in[V]},\{Y_{u}\}_{u\in [U]\setminus \{u'\}},\sum_{(u,v)\in[V]\times [U]}W_{u,v};\{W_{u',v}\}_{v\in\Vc}   \Bigg|\{W_{i,j},Z_{i,j}\}_{(i,j)\in \Tc} \right)\\
 \ge& |\Vc|L-\underbrace{I\left(\sum_{(u,v)\in[V]\times [U]}W_{u,v};\{W_{u',v}\}_{v\in\Vc}   \Bigg|\{W_{i,j},Z_{i,j}\}_{(i,j)\in \Tc} \right)}_{\overset{(\ref{ind})}{=}0  }\notag\\
 &  -I\left(\{X_{u',v}\}_{v\in[V]},\{Y_{u}\}_{u\in [U]\setminus \{u'\}};\{W_{u',v}\}_{v\in\Vc}   \Bigg|\sum_{(u,v)\in[V]\times [U]}W_{u,v},\{W_{i,j},Z_{i,j}\}_{(i,j)\in \Tc} \right)
\label{eq: step 2, proof of lemma Z|Z>=VL}\\
\ge &|\Vc|L-\underbrace{I\left(\{X_{u',v}\}_{v\in[V]},\{Y_{u}\}_{u\in [U]\setminus \{u'\}};\{W_{u,v}\}_{(u,v)\in[U]\times[V]}   \Bigg|\sum_{(u,v)\in[V]\times [U]}W_{u,v},\{W_{i,j},Z_{i,j}\}_{(i,j)\in \Tc} \right)}_{\overset{(\ref{security})}{=}0 }\\
 =&|\Vc|L,
\end{align}
where the second term in \eqref{eq: step 2, proof of lemma Z|Z>=VL} is zero due to the independence of the inputs $\{W_k\}_{k\in [K]}$.

Next, consider (\ref{lemma3zt}). Partition the set $\mathcal{T}$ into disjoint subsets as
$\mathcal{T} = \bigcup_{i\in [U-1]} \mathcal{T}_i$,
where $\mathcal{T}_i \cap \mathcal{T}_j = \emptyset$ for $i\neq j$, and
$\mathcal{T}_i \subseteq \{(u_i,v)\}_{v\in [V]}$ for each $i\in[U-1]$.
Then $|\mathcal{T}| = \sum_{i=1}^{U-1} |\mathcal{T}_i|$.

Applying the chain rule of entropy, we have
\begin{align}
H\left( \{Z_{u,v}\}_{(u,v)\in\mathcal{T}} \right)
&= \sum_{i=1}^{U-1}
H\left(\{Z_{u,v}\}_{(u,v)\in\mathcal{T}_i}
\;\big|\; \{Z_{u,v}\}_{(u,v)\in\mathcal{T}_j,\,j<i} \right) \\
&\overset{(\ref{lemma3zv})}{\geq} \sum_{i=1}^{U-1} |\mathcal{T}_i| L \label{lm3tt1}\\
&= |\mathcal{T}| L.
\end{align}

Here, (\ref{lm3tt1}) follows from the fact that each $\mathcal{T}_i$ consists of keys
associated with a fixed server $u_i$, i.e., $\mathcal{T}_i \subseteq \{(u_i,v)\}_{v\in [V]}$, 
and hence inequality (\ref{lemma3zv}) applies to each conditional entropy term.

\end{IEEEproof}

Equipped with the Lemmas, the converse bounds on the communication rates $R_X$, $R_Y$, the individual key rate $R_Z$, and the source key rate $R_{Z_\Sigma}$ follow immediately.

\subsubsection{Proof of $R_X \ge 1$}
For any $(u,v) \in [U] \times [V]$, we have
\begin{align}
L_X &= H(X_{u,v}) \notag\\
&\ge H\Big(X_{u,v} \,\big|\, \{W_{i,j},Z_{i,j}\}_{(i,j)\in [U]\times [V]\backslash \{(u,v)\}} \Big)
\overset{(\ref{lemma1X>=L})}{\ge} L,\\
\Rightarrow \quad R_X &= \frac{L_X}{L} \ge 1.
\end{align}

\subsubsection{Proof of $R_Y \ge 1$}
For any $u \in [U]$, we similarly have
\begin{align}
L_Y &= H(Y_u) \notag\\
&\ge H\Big(Y_u \,\big|\, \{W_{i,j},Z_{i,j}\}_{(i,j)\in [U]\times [V]\backslash \{(u,v)\}} \Big)
\overset{(\ref{lemma1Y>=L})}{\ge} L,\\
\Rightarrow \quad R_Y &= \frac{L_Y}{L} \ge 1.
\end{align}

Note that the communication rate constraints are independent of the security requirements. 
Since the server must recover the sum of all users' inputs, each individual input of size $L$ 
must be conveyed to the server. Accordingly, $R_X \ge 1$ corresponds to the minimum rate 
required on the user-to-server links, while $R_Y \ge 1$ corresponds to the minimum rate 
required on the server-to-server links.

\subsubsection{Proof of $R_Z \ge 1$} 
To ensure information-theoretic security, each user input must be concealed by randomness provided by its associated relay. 
By Shannon's perfect secrecy result \cite{shannon1949communication}, the entropy of each individual key cannot be smaller than that of the corresponding protected data, 
which implies $H(Z_{u,v}) \ge H(W_{u,v})$ for all $(u,v)\in[U]\times[V]$.

We first establish that the transmitted message $X_{u,v}$ reveals no information about the corresponding input $W_{u,v}$, i.e., $X_{u,v}$ is independent of $W_{u,v}$:
\begin{align}
I(X_{u,v};W_{u,v}) 
&\le I\Big(X_{u,v}, \sum_{(i,j)\in[V]\times [U]} W_{i,j}, \{W_{i,j}, Z_{i,j}\}_{(i,j)\in \Tc}; W_{u,v} \Big) \\
&= \underbrace{I\Big(\sum_{(i,j)\in[V]\times [U]} W_{i,j}, \{W_{i,j}, Z_{i,j}\}_{(i,j)\in \Tc}; W_{u,v} \Big)}_{\overset{(\ref{ind})}{=}0} \notag\\
&\quad + I\Big(X_{u,v}; W_{u,v} \,\Big|\, \sum_{(i,j)\in[V]\times [U]} W_{i,j}, \{W_{i,j}, Z_{i,j}\}_{(i,j)\in \Tc} \Big) \label{indikeyt1} \\
&\le \underbrace{I\Big(\{Y_k\}_{k\in[U]\setminus \{u\}}, \{X_{u,v}\}_{v\in [V]}; W_{[U]\times [V]} 
\,\Big|\, \sum_{(i,j)\in[V]\times [U]} W_{i,j}, \{W_{i,j}, Z_{i,j}\}_{(i,j)\in \Tc} \Big)}_{\overset{(\ref{security})}{=}0} \label{indikeyt2} \\
&= 0, \label{indikeyt3}
\end{align}
where the first term in (\ref{indikeyt1}) vanishes because the sum of inputs and the keys in $\Tc$ are independent of $W_{u,v}$, 
and (\ref{indikeyt2}) vanishes due to the security constraint.

Next, for any $(u,v)\in[U]\times[V]$, we have
\begin{align}
L_Z &= H(Z_{u,v}) \\
&\ge H(Z_{u,v}|W_{u,v}) \\
&\ge I(X_{u,v}; Z_{u,v} \,|\, W_{u,v}) \\
&= H(X_{u,v}|W_{u,v}) - \underbrace{H(X_{u,v}|W_{u,v}, Z_{u,v})}_{\overset{(\ref{messageX})}{=}0} \\
&= H(X_{u,v}) - \underbrace{I(X_{u,v}; W_{u,v})}_{\overset{(\ref{indikeyt3})}{=}0} \\
&\ge H\Big(X_{u,v} \,\Big|\, \{W_{i,j}, Z_{i,j}\}_{(i,j)\in [U]\times [V]\backslash \{(u,v)\}} \Big) \\
&\overset{(\ref{lemma1X>=L})}{\ge} L, \\
\Rightarrow \quad R_Z &= \frac{L_Z}{L} \ge 1.
\end{align}

This establishes that the individual key rate $R_Z$ must be at least $1$, 
which ensures that each input is perfectly concealed by the corresponding key.

\subsubsection{Proof of $R_{Z_{\Sigma}}\ge \min\{U+V+T-2,\, UV-1\}$}

First, observe that
\begin{align}
\min\{U+V+T-2,\, UV-1\} =
\begin{cases}
U+V+T-2, & \text{if } T \le (U-1)(V-1),\\
UV-1, & \text{if } T \ge (U-1)(V-1).
\end{cases}
\end{align}

When $T \le (U-1)(V-1)$, it suffices to show that $R_{Z_{\Sigma}} \ge U+V+T-2$. 
In particular, at $T=(U-1)(V-1)$, this bound reduces to 
$R_{Z_{\Sigma}} \ge U+V+(U-1)(V-1)-2 = UV-1$. 
Since increasing the number of colluding users $T$ cannot decrease the optimal source key rate, 
it follows that $R_{Z_{\Sigma}} \ge UV-1$ for all $T \ge (U-1)(V-1)$. 
Therefore, it is sufficient to establish the bound $R_{Z_{\Sigma}} \ge U+V+T-2$ under the condition $T \le (U-1)(V-1)$.

Recall that in the proof of the above Lemmas, the colluding set is chosen as 
$\mathcal{T} \subseteq ([U]\setminus \{u'\}) \times ([V]\setminus \{v'\})$, ensuring that $|\mathcal{T}| \le T \le (U-1)(V-1)$. 
We now select a colluding set $\mathcal{T}$ with $|\mathcal{T}|=T$ such that, for every server $u \in [U]\setminus\{u'\}$, 
there exists at least one user $(u,v_u) \in \mathcal{M}_u$ that does not belong to $\mathcal{T}$. 
Such a choice of $\mathcal{T}$ is feasible since $T \le (U-1)(V-1)$. 

With this choice, we have
\begin{align}
L_{Z_{\Sigma}} &= H(Z_{\Sigma}) \\
&\overset{(\ref{sourcekey})}{\ge} H(Z_{[U]\times [V]}) \\
&= H(\{Z_{u,v}\}_{(u,v)\in \mathcal{T}}) + H(Z_{[U]\times [V]} | \{Z_{u,v}\}_{(u,v)\in \mathcal{T}}) \\
&\overset{(\ref{lemma3zt})}{\ge} T L + H(Z_{[U]\times [V]} | \{W_{u,v}, Z_{u,v}\}_{(u,v)\in \mathcal{T}}) \label{sourcekeytt1} \\
&\ge T L + H(Z_{[U]\times [V]} | W_{[U]\times [V]}, \{W_{u,v}, Z_{u,v}\}_{(u,v)\in \mathcal{T}}) \\
&\ge T L + I\Big(Z_{[U]\times [V]}; \{Y_u\}_{u\in [U]\setminus\{u'\}}, \{X_{u',v}\}_{v\in [V]} \,\big|\, W_{[U]\times [V]}, \{W_{u,v}, Z_{u,v}\}_{(u,v)\in \mathcal{T}}\Big) \\
&\ge T L + H\Big(\{Y_u\}_{u\in [U]\setminus\{u'\}}, \{X_{u',v}\}_{v\in [V]} \,\big|\, W_{[U]\times [V]}, \{W_{u,v}, Z_{u,v}\}_{(u,v)\in \mathcal{T}}\Big) \notag\\
&\quad - \underbrace{H\Big(\{Y_u\}_{u\in [U]\setminus\{u'\}}, \{X_{u',v}\}_{v\in [V]} \,\big|\, Z_{[U]\times [V]}, W_{[U]\times [V]}, \{W_{u,v}, Z_{u,v}\}_{(u,v)\in \mathcal{T}}\Big)}_{\overset{(\ref{messageX})(\ref{messageY})}{=}0} \\
&= T L + H\Big(\{Y_u\}_{u\in [U]\setminus\{u'\}}, \{X_{u',v}\}_{v\in [V]} \,\big|\, \{W_{u,v}, Z_{u,v}\}_{(u,v)\in \mathcal{T}}\Big) \notag\\
&\quad - I\Big(\{Y_u\}_{u\in [U]\setminus\{u'\}}, \{X_{u',v}\}_{v\in [V]}; W_{[U]\times [V]} \,\big|\, \{W_{u,v}, Z_{u,v}\}_{(u,v)\in \mathcal{T}}\Big) \\
&= T L + H(\{X_{u',v}\}_{v\in [V]} | \{W_{u,v}, Z_{u,v}\}_{(u,v)\in \mathcal{T}}) 
+ H(\{Y_u\}_{u\in [U]\setminus\{u'\}} | \{X_{u',v}\}_{v\in [V]}, \{W_{u,v}, Z_{u,v}\}_{(u,v)\in \mathcal{T}}) \notag\\
&\quad - I\Big(\{Y_u\}_{u\in [U]\setminus\{u'\}}, \{X_{u',v}\}_{v\in [V]}, \sum_{(u,v)\in [U]\times [V]} W_{u,v}; W_{[U]\times [V]} \,\big|\, \{W_{u,v}, Z_{u,v}\}_{(u,v)\in \mathcal{T}}\Big) \\
&\overset{(\ref{lemma2xvwz}), (\ref{lemma2yuwz})}{\ge} T L + V L + (U-1)L 
- \underbrace{I\Big(\sum_{(u,v)\in [U]\times [V]} W_{u,v}; W_{[U]\times [V]} \,\big|\, \{W_{u,v}, Z_{u,v}\}_{(u,v)\in \mathcal{T}}\Big)}_{=L} \notag\\
&\quad - \underbrace{I\Big(\{Y_u\}_{u\in [U]\setminus\{u'\}}, \{X_{u',v}\}_{v\in [V]}; W_{[U]\times [V]} \,\big|\, \sum_{(u,v)\in [U]\times [V]} W_{u,v}, \{W_{u,v}, Z_{u,v}\}_{(u,v)\in \mathcal{T}}\Big)}_{\overset{(\ref{security})}{=}0} \label{sourcekeytt2} \\
&= T L + V L + (U-1)L - L \\
&= (V + U + T - 2)L.
\end{align}

Equation (\ref{sourcekeytt1}) follows from choosing $|\mathcal{T}|=T$ and applying (\ref{lemma3zt}). 
The fourth term in (\ref{sourcekeytt2}) equals $L$ because the sum conveys exactly one symbol of information about the inputs, which is independent of the keys held by the colluding users.

\section{Conclusion}

In this paper, we studied the problem of multi-server secure aggregation from an information-theoretic perspective. We considered a setting in which multiple aggregation servers jointly compute the sum of users' inputs, with each server connected to a group of users and servers exchanging messages over a fully connected network. Under an honest-but-curious threat model with user collusion, we characterized the fundamental limits of secure aggregation in terms of communication and key efficiency.

Our main result is a complete characterization of the optimal rate region for multi-server secure aggregation. The result precisely identifies the minimum user-to-server communication rate, server-to-server communication rate, individual key rate, and source key rate required to achieve correctness and perfect secrecy. The achievability is established via an explicit secure aggregation scheme, while the converse is proved using information-theoretic arguments based on entropy inequalities and security constraints.

The characterization of the optimal rate region reveals how the multi-server architecture fundamentally alters the tradeoff between security and efficiency compared to the typical single-server setting. In particular, it highlights the impact of inter-server communication and user collusion on the amount of key randomness required for secure aggregation. These results provide a theoretical benchmark for evaluating and designing communication- and key-efficient secure aggregation schemes in distributed learning systems.

Several directions for future work remain open. Extensions to partially connected server networks, dynamic user participation, and more general collusion models are of particular interest. Another promising direction is to investigate the implications of the derived information-theoretic limits for practical secure aggregation protocols in large-scale federated learning deployments.

\bibliographystyle{IEEEtran}
\bibliography{references_secagg.bib}
\end{document}

%% file: macros.tex
\long\def\comment#1{}

\newtheorem{example}{Example}
\newtheorem{theorem}{Theorem}
\newtheorem{lemma}{Lemma}
\makeatletter
\newcommand{\subalign}[1]{%
  \vcenter{%
    \Let@ \restore@math@cr \default@tag
    \baselineskip\fontdimen10 \scriptfont\tw@
    \advance\baselineskip\fontdimen12 \scriptfont\tw@
    \lineskip\thr@@\fontdimen8 \scriptfont\thr@@
    \lineskiplimit\lineskip
    \ialign{\hfil$\m@th\scriptstyle##$&$\m@th\scriptstyle{}##$\crcr
      #1\crcr
    }%
  }
}
\newcommand{\thickhline}{%
    \noalign {\ifnum 0=`}\fi \hrule height 1pt
    \futurelet \reserved@a \@xhline
}
\newcolumntype{"}{@{\hskip\tabcolsep\vrule width 1pt\hskip\tabcolsep}}

\makeatother

\newfont{\bbb}{msbm10 scaled 700}

\newfont{\bb}{msbm10 scaled 1100}

\def \rzsigmastar{{R_{Z_{\Sigma}}^*}}

\def  \zsigma{{Z_{\Sigma}}}
\def  \lzsigma{{L_{Z_{\Sigma}}}}

\newcommand{\msg}{message\xspace}

\newcommand{\Ip}{In particular\xspace}

\newcommand{\iid}{i.i.d.\xspace}

\newcommand{\info}{information\xspace}

\newcommand{\secagg}{secure aggregation\xspace}

\newcommand{\indep}{independent\xspace}

\newcommand{\indiv}{individual\xspace}

\newcommand{\hv}{{\bf h}}

\newcommand{\Rc}{{\cal R}}

\newcommand{\Tc}{{\cal T}}

\newcommand{\Vc}{{\cal V}}

\newcommand{\eqdef}{\stackrel{\Delta}{=}}

\newcommand{\be}{\begin{equation}}
\newcommand{\ee}{\end{equation}}
\newcommand{\bea}{\begin{eqnarray}}
\newcommand{\eea}{\end{eqnarray}}



%% file: author_TIT.tex
\author{
Zhou Li,~\IEEEmembership{Member,~IEEE},
Xiang~Zhang,~\IEEEmembership{Member,~IEEE}, 
Kai Wan,~\IEEEmembership{Member,~IEEE},
Hua Sun,~\IEEEmembership{Member,~IEEE},
Mingyue Ji,~\IEEEmembership{Member,~IEEE},
and Giuseppe Caire,~\IEEEmembership{Fellow,~IEEE}


\thanks{Z. Li is with the School of Computer, Electronics and Information, 
Guangxi University, Nanning 530004, China (e-mail: lizhou@gxu.edu.cn).}

\thanks{X. Zhang and G. Caire are with the Department of Electrical Engineering and Computer Science, Technical University of Berlin, 10623 Berlin, Germany (e-mail: \{xiang.zhang, caire\}@tu-berlin.de).
}

\thanks{
K. Wan is with the School of Electronic Information and Communications,
Huazhong University of Science and Technology, Wuhan 430074, China
(e-mail: kai\_wan@hust.edu.cn).}

\thanks{
H. Sun is with the Department of Electrical Engineering, University of North Texas, Denton, TX 76207 USA (e-mail: hua.sun@unt.edu).  }


\thanks{
M. Ji is with the Department of Electrical and Computer Engineering, University of Florida, Gainesville, FL 32611, USA
(e-mail: mingyueji@ufl.edu).    }

}

%% file: sections_v1/abstract.tex
\begin{abstract}
Secure aggregation is a fundamental primitive in privacy-preserving distributed learning systems, where an aggregator aims to compute the sum of users' inputs without revealing individual data. In this paper, we study a multi-server secure aggregation problem in a two-hop network consisting of multiple aggregation servers and multiple users per server, under the presence of user collusion. Each user communicates only with its associated server, while the servers exchange messages to jointly recover the global sum. We adopt an information-theoretic security framework, allowing up to $T$ users to collude with any server.

We characterize the complete optimal rate region in terms of user-to-server communication rate, server-to-server communication rate, individual key rate, and source key rate. Our main result shows that the minimum communication and individual key rates are all one symbol per input symbol, while the optimal source key rate is given by $\min\{U+V+T-2,\, UV-1\}$, where $U$ denotes the number of servers and $V$ the number of users per server. The achievability is established via a linear key construction that ensures correctness and security against colluding users, while the converse proof relies on tight entropy bounds derived from correctness and security constraints.

The results reveal a fundamental tradeoff between security and key efficiency and demonstrate that the multi-server architecture can significantly reduce the required key randomness compared to single-server secure aggregation. Our findings provide a complete information-theoretic characterization of secure aggregation in multi-server systems with user collusion.
\end{abstract}